\documentclass[10pt,journal]{IEEEtran}
\IEEEoverridecommandlockouts
\usepackage{cite}
\usepackage{color}
\usepackage{amsmath,amssymb,amsfonts, mathrsfs, bm, mathabx}
\usepackage{multirow}
\usepackage{algorithmic}
\usepackage{hyperref}
\usepackage{graphicx}
\usepackage{textcomp}
\usepackage{tabularx,booktabs}
\usepackage{stfloats}
\usepackage{longtable}
\usepackage[ruled]{algorithm2e}

\usepackage{url}
\usepackage{supertabular,booktabs}
\usepackage{array}
\usepackage{verbatim}
\usepackage{balance}

\def\BibTeX{{\rm B\kern-.05em{\sc i\kern-.025em b}\kern-.08em T\kern-.1667em\lower.7ex\hbox{E}\kern-.125emX}}

\newcommand {\mymarginpar}[1]{\marginpar{#1}}
\renewcommand {\marginpar}[1]{}
\def\_{\rule{.3em}{.15ex}}

\newcommand{\ls}[1]
{
    \dimen0=\fontdimen6\the\font
    \lineskip=#1\dimen0
    \advance\lineskip.5\fontdimen5\the\font
    \advance\lineskip-\dimen0
    \lineskiplimit=.9\lineskip
    \baselineskip=\lineskip
    \advance\baselineskip\dimen0
    \normallineskip\lineskip
    \normallineskiplimit\lineskiplimit
    \normalbaselineskip\baselineskip
    \ignorespaces
}

\newcommand {\bearn}{\begin{eqnarray*}}
\newcommand {\eearn}{\end{eqnarray*}}
\newcommand {\barr}{\begin{array}}
\newcommand {\earr}{\end{array}}

\newcommand {\N}{{\cal N}}

\newtheorem{definition}{Definition}
\newtheorem{property}[definition]{Property}
\newtheorem{proposition}[definition]{Proposition}
\newtheorem{lemma}[definition]{Lemma}
\newtheorem{theorem}[definition]{Theorem}
\newtheorem{corollary}[definition]{Corollary}
\newtheorem{example}{Example}
\newtheorem{remark}[definition]{Remark}

\newcommand {\benum} {\begin{enumerate}}
\newcommand {\eenum} {\end{enumerate}}
\newcommand {\bdesc} {\begin{description}}
\newcommand {\edesc} {\end{description}}

\newcommand {\bfig}[2] {
    \begin{figure}
    \centering
    \includegraphics[width=#2]{#1}
}
\newcommand {\brotatefig}[2]{
    \begin{figure}[htbp]
    \centerline{\epsfig{figure={#1},clip=,angle=-90,width={#2}}}
}
\newcommand {\bfigfirst}[2]{\begin{figure}[h]
                            \centerline {
                            \setlength{\epsfxsize}{#2}
                            \epsffile{#1}}}
\newcommand {\efig}[2]{\caption{#2}
                       \label{fig:#1}
                       \end{figure}
                       \mymarginpar{fig:#1}}
\newcommand {\erotatefig}[2]{\caption{#2}
                             \label{fig:#1}
                             \end{figure}
                             \mymarginpar{fig:#1}}
\newcommand {\rfig}[1]{Figure \ref{fig:#1}}

\newcommand {\btab}[1]{\begin{table}
                       \centering
                       \begin{tabular}{#1}}
\newcommand {\etab}[3]{\end{tabular}
                       \caption[#3]{#2}
                       \label{tab:#1}
                       \end{table}
                       \mymarginpar{tab:#1}
                       \vspace{.1in}}

\newcommand {\btabular}[1]{\begin{center}
                       \begin{tabular}{#1}}
\newcommand {\etabular}{\end{tabular}
                       \end{center}}

\newcommand {\bdefin}[1]{\begin{definition}
                      \mymarginpar{def:#1}
                      \label{def:#1} }
\newcommand {\edefin}{\end{definition}}
\newcommand {\rdef}[1]{Definition \ref{def:#1}}

\newcommand {\bpro}[1]{\begin{property}
                      \mymarginpar{pro:#1}
                      \label{pro:#1} }
\newcommand {\epro}   {\end{property}}

\newcommand {\bprop}[1]{\begin{proposition}
                      \mymarginpar{prop:#1}
                      \label{prop:#1} }
\newcommand {\eprop}{\end{proposition}}

\newcommand {\blem}[1]{\begin{lemma}
                      \mymarginpar{lem:#1}
                      \label{lem:#1} }
\newcommand {\elem}   {\end{lemma}}
\newcommand {\rlem}[1]{Lemma \ref{lem:#1}}

\newcommand {\bthe}[1]{\begin{theorem}
                      \mymarginpar{the:#1}
                      \label{the:#1} }
\newcommand {\ethe}   {\end{theorem}}
\newcommand {\rthe}[1]{Theorem \ref{the:#1}}

\newcommand {\bproof}{\noindent {\bf Proof.} \ }
\newcommand {\eproof} {\hfill \squares \\ \vspace{.3cm}}
\newcommand {\bcor}[1]{\begin{corollary}
                      \mymarginpar{cor:#1}
                      \label{cor:#1} }
\newcommand {\ecor}   {\end{corollary}}
\newcommand {\rcor}[1]{Corollary \ref{cor:#1}}

\newcommand {\bax}[1]{\begin{axiom}
                      \mymarginpar{ax:#1}
                      \label{ax:#1} }
\newcommand {\eax}       {\vspace{-.1in} \end{axiom}}

\newcommand {\bex}[2]{\vspace{.1in}
                      \begin{example}
                      \mymarginpar{ex:#1}
                       {\bf #2}
                      \label{ex:#1} }
\newcommand {\eex}       {\end{example} \vspace{.3cm} }
\newcommand {\rex}[1]{Example \ref{ex:#1}}

\newcommand {\brem}[1]{\begin{remark}
                      \mymarginpar{rem:#1}
                      \label{rem:#1} \em }
\newcommand {\erem}   {\end{remark}}

\newcommand {\beq}[1]{\mymarginpar{eq:#1}
                      \begin{equation}
                      \label{eq:#1} }

\newcommand {\beqno}[1]{\mymarginpar{eq:#1}
                      \begin{eqnarray}
                      \nonumber}

\newcommand {\eeq}       {\end{equation}}
\newcommand {\eeqno}       { && \end{eqnarray}}
\newcommand {\req}[1]{(\ref{eq:#1})}

\newcommand {\bear}[1]{\mymarginpar{eq:#1}
                       \begin{eqnarray}
                       \label{eq:#1} }

\newcommand {\bearno}[1]{\mymarginpar{eq:#1}
                       \begin{eqnarray}
                       \nonumber}

\newcommand {\eear}{\end{eqnarray}}
\newcommand {\eearno}{\end{eqnarray}}
\newcommand {\bsel}{\left \{ \begin{array}{cl}}
\newcommand {\esel}{\end{array} \right.}

\newcommand {\bmat}[1]{\left [ \begin{array}{#1}}
\newcommand {\emat}{\end{array} \right ]}
\newcommand {\bsec}[2]{\mymarginpar{sec:#2}
                       \section{#1}
                       \label{sec:#2} }

\newcommand {\rsec}[1]{Section \ref{sec:#1}}

\newcommand {\bsubsec}[2]{\mymarginpar{sec:#2}
                       \subsection{#1}
                       \label{sec:#2} }

\newcommand {\rsubsec}[1]{Section \ref{sec:#1}}

\def\R{I\kern-0.30em R}
\def\N{I\kern-0.30em N}
\def\P{I\kern-0.30em P}

\newcommand\squares{\vrule height6pt width7pt depth1pt}

\def\ex{{\bf\sf E}}
\def\pr{{\bf\sf P}}

\newcommand{\rhog}{\rho}

\newcommand{\trhoj}{\tilde \rho_j}
\newcommand{\trhojl}{\tilde \rho_{j,\ell}}
\newcommand{\trhojhl}{\tilde \rho_{j,\hat \ell}}
\newcommand{\Rj}{R_j}

\begin{document}

\title{Convolutional Coded Poisson Receivers}

\author{
    Cheng-En Lee, Kuo-Yu Liao, Hsiao-Wen Yu,  Ruhui Zhang, Cheng-Shang Chang,~\IEEEmembership{Fellow,~IEEE}, \\
    and Duan-Shin Lee,~\IEEEmembership{Senior Member,~IEEE}
    \thanks{
        The authors are with the Institute of Communications Engineering, National Tsing Hua University, Hsinchu  300044, Taiwan R.O.C. Email:
        benny\_110065508@gapp.nthu.edu.tw;
        d25602685@gmail.com;
        yuhw9817@gapp.nthu.edu.tw;
        huibrana@gapp.nthu.edu.tw;
	cschang@ee.nthu.edu.tw;
        lds@cs.nthu.edu.tw.

        Part of this work was presented in 2023 IEEE International Symposium on Information Theory (ISIT) \cite{upperboundISIT} and 2024 IEEE International Symposium on Information Theory (ISIT) \cite{potentialISIT}. This work was supported in part by the National Science and Technology, Taiwan, under Grant 111-2221-E-007-038-MY3, and in part by Qualcomm Technologies under Grant SOW NAT-487844-2.
        }
}

\maketitle

\begin{abstract}
In this paper, we present a framework for convolutional coded Poisson receivers (CCPRs) that incorporates spatially coupled methods into the architecture of coded Poisson receivers (CPRs). We use density evolution equations to track the packet decoding process with the successive interference cancellation (SIC) technique. We derive outer bounds for the stability region of CPRs when the underlying channel can be modeled by a $\phi$-ALOHA receiver. The stability region is the set of loads that every packet can be successfully received with a probability of 1. Our outer bounds extend those of the spatially-coupled Irregular Repetition Slotted ALOHA (IRSA) protocol and apply to channel models with multiple traffic classes. For CCPRs with a single class of users, the stability region is reduced to an interval. Therefore, it can be characterized by a percolation threshold. We study the potential threshold by the potential function of the base CPR used for constructing a CCPR. In addition, we prove that the CCPR is stable under a technical condition for the window size. For the multiclass scenario, we recursively evaluate the density evolution equations to determine the boundaries of the stability region. Numerical results demonstrate that the stability region of CCPRs can be enlarged compared to that of CPRs by leveraging the spatially-coupled method. Moreover, the stability region of CCPRs is close to our outer bounds when the window size is large.
\end{abstract}


\bsec{Introduction}{introduction}

The fifth-generation networks (5G) and beyond aim to support various connectivity classes of users that have different quality-of-service (QoS)  requirements, including (i) enhanced mobile broadband (eMBB), (ii) ultra-reliable low-latency communications (URLLC), and (iii) massive machine-type communications (mMTC) (see, e.g., \cite{li20175g,bennis2018ultra,popovski2019wireless,le2020overview} and references therein).  {\em Network slicing} \cite{popovski20185g} that partitions and allocates available radio resources to provide differentiated QoS  in a {\em single} radio network has received much attention lately. For downlink transmissions, network slicing could be tackled by centralized scheduling algorithms \cite{anand2020joint}. On the other hand, various schemes of coded multiple access (see, e.g., \cite{casini2007contention,liva2011graph,narayanan2012iterative,paolini2012random,paolini2015coded,jakovetic2015cooperative,sun2017coded,stefanovic2018coded,Hoshyar2008,SCMA,Yuan2016,Chen2017,ordentlich2017low,vem2019user,andreev2020polar}) have been proposed in the literature to address the problem of uplink transmissions. The framework of {\em Poisson receivers} \cite{chang2020Poisson} is an abstraction of these coded multiple access schemes to hide the encoding/decoding complexity of the physical layer from the Medium Access Control (MAC) layer.

A Poisson receiver specifies the probability that a packet is successfully received (decoded) when the number of packets transmitted simultaneously to the receiver follows a Poisson distribution (Poisson offered load). Inspired by  Irregular Repetition Slotted ALOHA (IRSA) \cite{liva2011graph} and coded slotted ALOHA (CSA) \cite{narayanan2012iterative,paolini2012random,paolini2015coded,jakovetic2015cooperative,sun2017coded,stefanovic2018coded}, a Poisson receiver can be used as a building block to construct a system of coded Poisson receivers (CPR) with multiple classes of users and multiple classes of receivers. In \cite{chang2022stability}, the stability region of a system of CPR is defined as the region of Poisson offered loads in which all the packets can be successfully received (with probability 1). The stability regions of various network slicing policies, including the complete sharing and the receiver reservation policies, were computed numerically in \cite{chang2022stability}. The numerical results in \cite{chang2022stability} showed that different network slicing policies lead to different stability regions. Given a targeted stability region in a network with various classes of traffic, one engineering problem is to design a system of CPR associated with a network slicing policy to cover the targeted stability region.

One objective of this paper is to explore the boundaries of the stability region of a system of CPRs and to devise schemes to approach the boundaries. The contributions of the paper are as follows:

\noindent (i) {\em Outer bounds for the stability region} (see \rthe{capacity}): For IRSA and CSA, there is only a single class of users and a single class of receivers. Thus, the stability region is reduced to an interval and its boundary is limited to one packet per time slot (as the underlying channel is the slotted ALOHA channel). When the underlying channel is replaced by the $D$-fold ALOHA channel (in which, at most, $D$ packets can be successfully received in a time slot), the boundary is limited to $D$ packets per time slot. These boundaries are simply the capacities of the underlying channels. As an application of the Area Theorem \cite{ashikhmin2004extrinsic}, tighter bounds for the stability regions for IRSA and coded $D$-fold ALOHA were derived in \cite{paolini2015coded} and \cite{stefanovic2017asymptotic}, respectively. In \rthe{capacity}, we derive outer bounds for the stability region of a system of CPR when the underlying channel is modeled by a $\phi$-ALOHA receiver \cite{liu2021aloha}. Such a model is a generalization of both the slotted ALOHA channel and the $D$-fold ALOHA channel. As such, our outer bounds recover the upper bounds for IRSA and coded $D$-fold ALOHA in \cite{paolini2015coded,stefanovic2017asymptotic} as special cases. Our outer bounds can also be applied to the setting with multiple traffic classes.

\noindent (ii) {\em Threshold saturation of Convolutional Coded Poisson Receivers} (see \rthe{saturation}): Motivated by the remarkable performance of convolutional Low-Density Parity-Check (LDPC) codes (see, e.g., \cite{kudekar2011threshold,yedla2012simple,mitchell2015spatially}) and the spatially coupled IRSA \cite{Liva2012spatially,stefanovic2017asymptotic},  we propose using the construction method for convolutional LDPC codes and spatially coupled IRSA to construct Convolutional Coded Poisson Receivers (CCPRs) with multiple classes of users and multiple classes of receivers. To construct a CCPR, one first constructs  $L$ stages of random bipartite graphs from a base CPR.  The $L$ stages of bipartite graphs are concatenated to form a single bipartite graph.  Then, we rewire the receiver ends of the $\ell$-th copy, where $1\le \ell\le L$, to a randomly selected copy in a window $w$. We refer the reader to \rsubsec{CCPRc} for more details on the constructions. By using the {\em puncturing} technique in LDPC codes, we show in \rthe{saturation} that the stability region with $L$ stages is not larger than that with $L-1$ stages, and thus the stability region converges as $L \to \infty$. Such a phenomenon is analogous to {\em threshold saturation} for convolutional LDPC codes and spatially coupled IRSA (see, e.g., \cite{kudekar2011threshold,Liva2012spatially,mitchell2015spatially,xu2021some}).

\noindent (iii) {\em Potential functions and percolation thresholds for CCPRs with a single class of users (see \rthe{CCPR_potential} and \rthe{saturation_singleclass})}: Even though we know that the stability region converges as $L \to \infty$, characterizing such a stability region is difficult for CCPR with multiple classes of users. For CCPRs with a single class of users, we use a remarkable mathematical tool called {\em potential functions} by Yedle et al. \cite{yedla2012simple, Yedla2012vector} and Schlegel et al. \cite{Schlegel2013Lyapunov} in their study of LDPC codes. In \rthe{CCPR_potential}, we derive the potential function of a (base) CPR. Based on such a potential function, we define three thresholds: the single-system threshold $G_s^*$, the potential threshold $G_{conv}^*$, and the potential bound $G_{up}^*$. We show the threshold saturation theorem in \rthe{saturation_singleclass} that for all $G \le G_{conv}^*$, a CCPR with $L$ stages is stable if the window size $w$ satisfies a technical condition. The three thresholds are related as follows:
\beq{main0123}
    G_s^* < G_{conv}^* < G_{up}^*.
\eeq
The single-system threshold $G_s^*$ recovers the percolation threshold of a (base) CPR in \cite{chang2022stability}. When the underlying channel is modeled by the $D$-fold ALOHA, the potential bound $G_{up}^*$ is the same as the upper bound for stability in \cite{stefanovic2017asymptotic}.

\noindent (iv) {\em Numerical results (see \rsec{numerical})}: We provide numerical results for various systems of convolutional coded Poisson receivers. For CCPRs with one class of users and one class of receivers, our numerical results for $G_{conv}^*$ and $G_{up}^*$ in various parameter settings match very well with those in Table 1 of \cite{Liva2012spatially} and Table I of \cite{stefanovic2017asymptotic}. We also study the convolutional IRSA with two classes of users and two classes of receivers. We consider the two network slicing policies in \cite{chang2022stability}: the complete sharing policy and the receiver reservation policy. For $L=40$, our numerical results show that the convolutional effect (due to spatial coupling) can indeed enlarge the stability region. Such enlargement appears to be monotone in the window size $w$.

The structure of this paper is organized as follows: In \rsec{poisson}, we provide a review of the coded Poisson receiver framework. In \rsec{upper}, we show the outer bounds of the stability region of CPRs. In \rsec{CCPR}, we introduce the concept of CCPRs and present the development of density evolution equations for their decoding process, as well as the threshold saturation phenomena. In \rsec{saturation}, we delve into the application of the potential function to our CCPR framework, focusing on a single class of users. We utilize potential functions to characterize three significant thresholds. We establish the saturation theorem in this section as well. Some numerical results for both single and multiple traffic classes are presented in \rsec{numerical}. We conclude the paper and discuss potential extensions of our work in \rsec{con}. Several proofs of theorems are included in Appendices \ref{proof_of_maxdecode}, \ref{proof_of_mainextc}, and \ref{proof_of_saturation_singleclass} of the supplemental material. We provide a list of notations in Appendix \ref{notation_list}.

\bsec{Review of the framework of coded Poisson receivers}{poisson}

For the paper to be self-contained, we briefly review the framework of coded Poisson receivers in \cite{chang2020Poisson,liu2021aloha,chang2022stability}.

\bsubsec{Poisson receivers and ALOHA receivers}{poissondef}

A Poisson receiver is an {\em abstract} receiver proposed in \cite{chang2020Poisson}.  The key insight of a Poisson receiver is to specify the packet success probability when the system is subject to Poisson arrivals. Specifically, a system with $K$ classes of input traffic is said to have a Poisson offered load  $\rho=(\rho_1, \rho_2, \ldots, \rho_K)$ if these $K$ classes of input traffic are {\em independent}, and the number of class $k$ packets arriving at the system follows a Poisson distribution with mean $\rho_k$, for $k=1,2, \ldots, K$.

\bdefin{Poissonmul}{(\bf Poisson receiver with multiple classes of input traffic \cite{chang2020Poisson})}
    An abstract receiver is called a {\em $(P_{{\rm suc},1}(\rhog), P_{{\rm suc},2}(\rhog), \ldots, P_{{\rm suc},K}(\rhog))$-Poisson receiver} with $K$ classes of input traffic if the receiver is subject to a Poisson offered load  $\rho=(\rho_1, \rho_2, \ldots, \rho_K)$, a tagged (randomly selected) class $k$ packet is successfully received with probability $P_{{\rm suc},k}(\rhog)$, for $k=1,2, \ldots, K$.
\edefin

The throughput of class $k$ packets (defined as the expected number of class $k$ packets that are successfully received) for a {\em $(P_{{\rm suc},1}(\rhog), P_{{\rm suc},2}(\rhog), \ldots, P_{{\rm suc},K}(\rhog))$-Poisson receiver} subject to a Poisson offered load $\rho$ is thus
\beq{Poithrmul}
    \Theta_k=\rho_k \cdot P_{{\rm suc},k}(\rhog),
\eeq
$k=1,2, \ldots, K$.

By viewing each time slot as a Poisson receiver, various well-known systems can be modeled by Poisson receivers in \cite{chang2020Poisson,liu2021aloha,chang2022stability}. These include Slotted ALOHA (SA) \cite{ALOHA}, SA with multiple cooperative receivers, and Rayleigh block fading channel with capture \cite{stefanovic2014exploiting,clazzer2017irregular,stefanovic2018coded}. It can also be applied to the setting with multiple packet reception in a time slot (see, e.g., \cite{ghanbarinejad2013irregular,ordentlich2017low,stefanovic2017asymptotic,glebov2019achievability,vem2019user,chen2022analytic}).

Like a Poisson receiver, an ALOHA receiver \cite{liu2021aloha} is also an abstract receiver. Such an abstract receiver treats the physical layer with a {\em deterministic} input-output function. Denote by ${\cal Z}^+$ the set of nonnegative integers. We say a system with $K$ classes of input traffic is subject to a {\em deterministic} load   $n=(n_1, n_2, \ldots, n_K)  \in {{\cal Z}^+}^K$ if the number of class $k$ packets arriving at the system is $n_k$.

\bdefin{phiALOHA}{\bf (ALOHA receiver with multiple classes of input traffic \cite{liu2021aloha})}
    Consider a  deterministic function $$\phi: {{\cal Z}^+}^K \rightarrow {{\cal Z}^+}^K$$ that maps a $K$-vector $n=(n_1, n_2, \ldots, n_K)$ to the $K$-vector $(\phi_1(n), \phi_2(n), \ldots, \phi_K(n))$. An abstract receiver is called a $\phi$-ALOHA receiver (with $K$ classes of input traffic) if the number of class $k$ packets that are successfully received is exactly $\phi_k(n)$, $k=1,2, \ldots, K$, when the receiver is subject to a deterministic load $n=(n_1, n_2, \ldots, n_K)$.
\edefin

One typical example of the $\phi$-ALOHA receiver is a time slot in the slotted ALOHA (SA) system, where at most one packet can be received. The $D$-fold ALOHA system proposed in \cite{ordentlich2017low} is a generalization of the SA system. If there are less than or equal to  $D$ packets transmitted in a time slot, then all these packets can be successfully received. On the other hand, if there are more than $D$ packets transmitted in a time slot, then all these packets are lost. Thus, a time slot in the $D$-fold ALOHA system is a $\phi$-ALOHA receiver (with a single class of input traffic), where
\beq{phi7722}
    \phi(n)=\left \{\begin{array}{cc}
        n & \mbox{if}\; n \le D \\
        0 & \mbox{otherwise}
    \end{array} \right ..
\eeq

Another example of a $\phi$-ALOHA receiver is the model for the near-far \cite{ordentlich2017low} decoding scheme. In such a model, there are two classes of input traffic and the power from one class (the near users) is much stronger than that of the other class (the far users). By decoding the near users first and using the SIC technique, a time slot in the near-far SIC decoding scheme can be modeled by a $\phi$-ALOHA with
\beq{phi7724}
    \phi(n_1,n_2)=\left \{\begin{array}{cc}
        (n_1,n_2) & \mbox{if}\; (n_1,n_2) \le (1,1) \\
        (0,0) & \mbox{otherwise}
    \end{array} \right ..
\eeq

It was shown in Theorem 14 of \cite{liu2021aloha} that for every ALOHA receiver, there is an induced Poisson receiver. This is done by computing the throughput of the $\phi$-ALOHA receiver when it is subject to the Poisson offered load  $\rho$ and then using \req{Poithrmul} to find the success probability function. For the $D$-fold ALOHA system, the throughput for a Poisson offered load $\rho$ is
$$ \sum_{t=1}^{D} t\frac{e^{-\rho} \rho^{t}}{t!}=\rho \sum_{t=0}^{D-1} \frac{e^{-\rho} \rho^{t}}{t!}.$$
From \req{Poithrmul}, it is a Poisson receiver with the following success probability function:
\beq{tfold1111}
    P_{\rm suc}(\rhog)=\sum_{t=0}^{D-1} \frac{e^{-\rho} \rho^{t}}{t!}.
\eeq

\bsubsec{Coded Poisson receivers with multiple classes of users and receivers}{mulr}

The idea of using the SIC technique in the Contention Resolution Diversity Slotted ALOHA (CRDSA) protocol \cite{casini2007contention} and the Irregular Repetition Slotted ALOHA (IRSA) protocol leads to the development of coded Poisson receivers (CPR) with multiple classes of users and receivers in \cite{chang2022stability}. As in \cite{chang2022stability}, let us consider a system with $G_{k} T$ class $k$  users, $k=1,2, \ldots, K$, and $F_j T$ class $j$ Poisson receivers, $j=1,2, \ldots, J$. Class $j$ Poisson receivers have the success probability functions $P_{{\rm suc},1,j}(\rhog), P_{{\rm suc},2,j}(\rhog), \ldots, P_{{\rm suc},K,j}(\rhog)$ for the $K$ classes of input traffic. Each class $k$ user transmits its packet  for (a random number of) $L_k \ge 1$ times (copies). With the routing probability $r_{k,j}$ ($\sum_{j=1}^J r_{k,j}=1$), each copy of a class $k$ packet is transmitted  {\em uniformly} and {\em independently} to one of the $F_j T$ class $j$ Poisson receivers.

There are two assumptions made in \cite{chang2022stability}:
\begin{description}
    \item[(i)] Perfect SIC: as long as one copy of a packet is successfully received by one of the receivers, then it can be used to remove the other copies of that packet from the other receivers.
    \item[(ii)] Independent Poisson receivers: the event that a packet is successfully received by a Poisson receiver is independent of the outcomes of the other Poisson receivers as long as their input traffic is independent of each other.
\end{description}

As in \cite{liva2011graph,chang2022stability}, let $\Lambda_{k,d}$ be the probability that a class $k$ packet is transmitted $d$ times, i.e.,
\beq{extm1111}
    P(L_{k}=d)=\Lambda_{k,d}, \;d=1,2,\ldots
\eeq
Define the generating function
\beq{mean0000mulr}
    \Lambda_{k}(x)=\sum_{d=0}^\infty \Lambda_{k,d} \cdot x^d
\eeq
of the {\em degree distribution} of a class $k$ user node, and the generating function
\beq{mean3333mulr}
    \lambda_{k}(x)=\sum_{d=0}^\infty \lambda_{k,d} \cdot x^d
\eeq
of the {\em excess degree distribution} of a class $k$ user node, where
\beq{mean2222mulr}
    \lambda_{k,d}=\frac{\Lambda_{k,d+1}\cdot (d+1)}{\sum_{d=0}^\infty\Lambda_{k,d+1}\cdot (d+1)}
\eeq
is the probability that the user end of a randomly selected class $k$ edge has additional $d$ edges, excluding the randomly selected class $k$ edge. Note that the mean degree of a class $k$ user node is
\beq{mean1111mulr}
    \Lambda_{k}^{\prime}(1)=\sum_{d=0}^\infty d \cdot \Lambda_{k,d},
\eeq
and that
\beq{mean3344mulr}
    \lambda_{k}(x)=\frac{\Lambda_{k}^\prime(x)}{\Lambda_{k}^{\prime}(1)}.
\eeq

Let $$G=(G_1, G_2, \ldots, G_{K}),$$
$$\Lambda^\prime (x)=(\Lambda^\prime_1 (x), \Lambda^\prime_2 (x), \ldots, \Lambda^\prime_K (x)),$$
\beq{mulm4444}
    \Rj=(\frac{r_{1,j}}{F_j}, \frac{r_{2,j}}{F_j}, \ldots, \frac{r_{K,j}}{F_j}),
\eeq
and
\beq{mean6666dmulr}
    \trhoj= G \circ \Lambda^\prime(1)\circ \Rj,
\eeq
where $\circ$ denotes the element-wise multiplication of two vectors.

Using the tree evaluation (density evolution) method in \cite{luby1998analysis,liva2011graph,paolini2011graph,paolini2012random} and the reduced Poisson offered load argument in \cite{chang2020Poisson}, the following result was derived in Theorem 1 of \cite{chang2020Poisson}.

\bthe{mainext}(Theorem 1 of \cite{chang2020Poisson})
    As $T \to \infty$, the system of CPRs after the $i$-th SIC iteration converges to a Poisson receiver with the success probability function for the class $k$ traffic
    \bear{mean8888thumulam}
        \tilde P_{{\rm suc},k}^{(i)}(G) =1-\Lambda_k \Big  (1- \sum_{j=1}^J r_{k,j} P_{{\rm suc},k,j}(q^{(i-1)} \circ \trhoj)\Big),
    \eear
    $k=1,2, \ldots, K$, where $q^{(i)}=(q_{1}^{(i)}, q_{2}^{(i)}, \ldots, q_{K}^{(i)})$ can be computed recursively from the following equation:
    \beq{tag6666dmulm}
        q_{k}^{(i+1)}=\lambda_{k}\Big (1-\sum_{j=1}^J r_{k,j} P_{{\rm suc},k,j}(q^{(i)} \circ \trhoj)\Big ),
    \eeq
    with $q^{(0)}=(1, 1, \ldots, 1)$.
\ethe

If the number of copies $L_k \ge 2$ for all $k$, i.e, $\Lambda_{k,1}=0$,  then the generating function $\lambda_k(x)$ is strictly increasing for $0 \le x \le 1$, and thus invertible. By letting $p_{k}^{(i+1)}= \lambda_k^{-1}(q_{k}^{(i+1)})$, one can rewrite \req{tag6666dmulm} as follows:
\beq{key1111}
    p_{k}^{(i+1)}=1-\sum_{j=1}^J r_{k,j} P_{{\rm suc},k,j}(G \circ \Lambda^\prime(p^{(i)}) \circ \Rj),
\eeq
where $p^{(i)}=(p_1^{(i)}, p_2^{(i)}, \ldots, p_K^{(i)})$ and $p^{(0)}=(1, 1, \ldots, 1)$.

Note that a system of CPRs with $K$ classes of users and $J$ classes of receivers is characterized by the following parameters: (i) the total number of Poisson receivers $T$, (ii) the offered load vector $G=(G_1,G_2, \ldots, G_K)$, (iii) the degree distribution vector $\Lambda(x)=(\Lambda_1(x),\Lambda_2(x), \ldots, \Lambda_K(x))$, (iv) the $K \times J$ routing matrix $R=(r_{k,j})$, and (v) the partition vector $F=(F_1, F_2, \ldots, F_J)$.  In \cite{chang2022stability}, the routing matrix and the partition vector correspond to network slicing policies. To ease the presentation, we denote a system of CPRs with $K$ classes of users and $J$ classes of receivers as the $(T,G, \Lambda(x), R, F)$-CPR.

\bsubsec{Stability}{stability}

The stability of a system of coded Poisson receivers in \cite{chang2022stability} is motivated by the stability of a queueing system, where the system is (rate) stable if the departure rate is the same as the arrival rate. In that regard, a system of coded Poisson receivers is stable if all its packets can be successfully received. The precise definition is given below.

\bdefin{srates}{(\bf Stability of coded Poisson receivers with multiple classes of input traffic \cite{chang2022stability})}
    Consider the $(T,G, \Lambda(x), R, F)$-CPR described in \rsec{mulr}. A Poisson offered load $G=(G_1, G_2, \ldots, G_K)$ to a system of coded Poisson receivers is said to be {\em stable} if, as $T \to \infty$, the probability that a packet is successfully received approaches 1 when the number of iterations goes to infinity, i.e.,
    \beq{srates1111}
        \lim_{i \to\infty} \tilde P_{{\rm suc},k}^{(i)}(G)=1, \quad k=1,2,\ldots, K,
    \eeq
    where $\tilde P_{{\rm suc},k}^{(i)}(G)$ is defined in \req{mean8888thumulam}.
\edefin

As in \cite{chang2022stability}, we make the following four assumptions for the stability analysis:

\begin{description}
    \item[(A1)] For all $k=1, \ldots, K$ and $j=1,2, \ldots, J$, the success probability function $P_{{\rm suc},k,j}(\rho)$ is a continuous and decreasing function of $\rho$, and $P_{{\rm suc},k,j}({\bf 0})=1$, where ${\bf 0}$ is the vector with all its elements being 0. Furthermore, for the analysis in \rsec{saturation}, we require $P_{{\rm suc},k,j}(\rho)$ to have a continuous second derivative.
    \item[(A2)] If $\rho \ne {\bf 0}$, then $P_{{\rm suc},k,j}(\rho)<1$ for all $k=1, \ldots, K$, and $j=1,2, \ldots, J$.
    \item[(A3)] $r_{k,j}>0$ for all $k=1, \ldots, K$, and $j=1,2, \ldots, J$.
    \item[(A4)] Every packet is transmitted at least twice, i.e., $\Lambda_{k,1}=0$ for all $k=1,2, \ldots, K$.
\end{description}

Under these assumptions, a necessary and sufficient condition for a  Poisson offered load $G$ to be stable is presented in \rthe{stable}. A monotonicity result is presented in \rthe{region}. These two theorems appear in Theorems 2 and 3 in \cite{chang2022stability}. They are duplicated here for their importance and for the completeness of this paper.
\bthe{stable}(Theorem 2  of \cite{chang2022stability})
    Under (A1), a Poisson offered load $G$ is stable if $q={\bf 0}$ is the unique solution  in $[0,1]^K$ of the following  $K$ equations:
    \beq{stable0000}
        q_{k}=\lambda_{k}\Big (1- \sum_{j=1}^J r_{k,j} P_{{\rm suc},k,j}(q \circ \trhoj)\Big ),
    \eeq
    $k=1,2, \ldots, K$, where $\trhoj$ is defined in \req{mean6666dmulr} and $q=(q_1,q_2,\ldots,q_K)$. On the other hand, under (A1), (A2), and (A3), a positive Poisson offered load $G$ (with $G_k > 0$ for all $k$) is stable only if $q={\bf 0}$ is the unique solution in $[0,1]^K$ of the $K$ equations in \req{stable0000}. Moreover, by \req{key1111}, under (A1), (A2), (A3) and (A4), a positive  Poisson offered load $G$ (with $G_k > 0$ for all $k$) is stable if and only if $p={\bf 0}$ is the unique solution  in $[0,1]^K$ of the following $K$ equations:
    \beq{key2222}
        p_k=1-\sum_{j=1}^J r_{k,j} P_{{\rm suc},k,j}(G \circ \Lambda^\prime(p) \circ \Rj),
    \eeq
    $k=1,2, \ldots, K$ and $p=(p_1,p_2,\ldots,p_K)$.
\ethe

\bthe{region}(Theorem 3 in \cite{chang2022stability})
        Suppose that (A1), (A2), and (A3) hold. If a positive Poisson offered load $\hat G=(\hat G_1, \hat G_2, \ldots, \hat G_K)$ (with $\hat G_k >0$ for all $k$) is stable, then any Poisson offered load $G$ with $G \le \hat G$ is also stable.
\ethe

The monotonicity results in \rthe{region} leads to the notion of the stability region in \cite{chang2022stability}.

\bdefin{region}
    Under (A1), (A2), and (A3), the stability region $S$ is defined as the maximal stable set such that (i) any $G \in S$ is stable, and (ii) any $G \not \in S$ is not stable.
\edefin

\bsec{Outer bounds for the stability region}{upper}

In this section, we derive an outer bound for the stability region for a system of coded Poisson receivers when the Poisson receivers are induced from $\phi$-ALOHA receivers. For this, we need to define the concept of ``capacity'' for $\phi$-ALOHA receivers. Analogous to the terminology used in queueing theory, we use ``capacity'' for systems with deterministic loads and ``stability'' for systems with stochastic loads.

For a $\phi$-ALOHA receiver, the {\em failure} function $\phi^c$ \cite{liu2021aloha} is defined by
\beq{phi0012}
    \phi^c(n)= n-\phi(n).
\eeq
A $\phi$-ALOHA receiver  is called {\em monotone} if the {\em failure} function $\phi^c$ is {\em increasing} in the deterministic load $n$, i.e., for any $n^{\prime} \le n^{\prime\prime}$,
\beq{phi0000}
    \phi^c(n^{\prime}) \le \phi^c(n^{\prime\prime}).
\eeq
The failure function represents the number of packets remaining to be decoded. With the monotonicity, we now define the capacity region.

\bdefin{capacity}{(\bf Capacity of a $\phi$-ALOHA receiver)}
    For a monotone $\phi$-ALOHA receiver, its capacity region $S$ is defined to be the set of deterministic loads such that all the packets are successfully received, i.e.,
    \beq{capacity1111}
        S=\{n: \phi(n)=n\}.
    \eeq
    An $K$-vector $(b_1, b_2, \ldots, b_K)$ is called an {\em affine capacity envelope} with the bound $B$ if for  $n \in S$,
    \beq{env1111}
        \sum_{k=1}^K b_k \phi_k(n) \le B.
    \eeq
\edefin

For instance, for the $D$-fold ALOHA system with $K$ classes of users, it has the capacity region $S=\{n: \sum_{k=1}^K n_k \le D\}$ as the total number of packets that can be successfully received in a time slot is at most $D$. As such, the $K$-vector $(1,1,\ldots, 1)$ is also an affine capacity envelope with the bound $D$ for the $D$-fold ALOHA system with $K$ classes of users.

We will use affine capacity envelopes to derive outer bounds for the stability region for a system of CPRs. For this, we need two properties of $\phi$-ALOHA receivers in \cite{liu2021aloha}: the closure property and the all-or-nothing property. As the failure function represents the number of packets that remain to be decoded, we can decode the remaining packets for the second time by removing those successfully decoded packets. This corresponds to the SIC technique in the literature. The number of packets remaining to be decoded after the second time of decoding is $ \phi^c (\phi^c (n))$. Intuitively, we can carry out the iterative decoding approach (for an infinite number of times) until no more packets can be decoded. A $\phi$-ALOHA receiver that always does the iterative decoding approach until no more packets can be decoded is said to satisfy the {\em closure} property, i.e.,
\beq{phiclosed}
    \phi^c (\phi^c (n))= \phi^c(n),
\eeq
for all $n$. In addition to this, we also need the all-or-nothing property. A $\phi$-ALOHA receiver with $K$ classes of input traffic is said to be an {\em all-or-nothing} receiver if it is {\em monotone} and satisfies the {\em all-or-nothing property}, i.e.,  either $\phi_k(n)=n_k$ or $\phi_k(n)=0$ for all $n=(n_1, n_2, \ldots,n_K)$ and $k=1,2,\ldots, K$. The all-or-nothing property implies {\em if class $k$ packets (for some $k$) are successfully decoded, then they are all decoded during the same iteration}.

In the following theorem, we derive an outer bound for the stability region of a system of coded Poisson receivers  by using the concept of the affine capacity envelope.

\bthe{capacity}
    Consider the $(T,G, \Lambda(x), R, F)$-CPR with $T$ Poisson receivers being induced from a $\phi$-ALOHA receiver. Suppose that the $\phi$-ALOHA receiver  that satisfies the closure property and the all-or-nothing property, and it has an affine capacity envelope $(b_1, b_2, \ldots, b_K)$ with the bound $B$ and that $b_k$'s are binary. Then for any stable offered load $G$,
    \bear{env4444}
        &&\sum_{k=1}^K b_k G_k \nonumber\\
        &&\le \sum_{j=1}^J F_j \Big (\sum_{\tau=0}^{B-1} \tau e^{-\mu_j} \frac{{\mu_j}^\tau}{\tau!}+B (1-\sum_{\tau=0}^{B-1}e^{-\mu_j} \frac{{\mu_j}^\tau}{\tau!}) \Big ), \nonumber\\
    \eear
    where
    \beq{env6666}
        \mu_j =\sum_{k=1}^K b_k G_k \Lambda_k^\prime(1) r_{k,j}/F_j.
    \eeq
\ethe

For the proof of \rthe{capacity}, we need the following lemma. The proof is presented in Appendix \ref{proof_of_maxdecode}.

\blem{maxdecode}
Under the assumptions in \rthe{capacity}, the number of packets that are {\em actually} decoded by any one of the $T$ Poisson receivers is within the capacity region of the $\phi$-ALOHA receiver.
\elem

\bproof(\rthe{capacity})
Let $X_k(t)$ be the number of class $k$ packets sent to the $t$-th Poisson receiver and $Y_k(t)$ be the number of class $k$ packets that are {\em actually} decoded by the $t$-th Poisson receiver. In view of \rlem{maxdecode} and the definition of the affine capacity envelope in \req{env1111}, we have
\beq{env2211}
    \sum_{k=1}^K b_k Y_k(t) \le B.
\eeq

Note that \req{env2211} holds trivially when nothing is decoded in the $t$-th Poisson receiver, i.e., $Y_k(t)=0$ for all $k$. Since $Y_k(t) \le X_k(t)$, we have from \req{env2211} that
\beq{env2222}
    \sum_{k=1}^K b_k Y_k(t) \le \min[\sum_{k=1}^K b_k X_k(t), B].
\eeq

Taking expectations on both sides of \req{env2222} yields
\beq{env2233}
    \sum_{k=1}^K b_k \ex[Y_k(t)] \le \ex[\min[\sum_{k=1}^K b_k X_k(t), B]].
\eeq

If the $t$-th Poisson  receiver is a class $j$ receiver, then we have from \req{mean6666dmulr} that $X_k(t)$,
$k=1,2, \ldots, K$, are independent Poisson random variables with mean
\beq{env5555}
    \rho_{k,j}=G_k \Lambda_k^\prime(1) r_{k,j}/F_j.
\eeq

Since $b_k's$ are binary and the sum of independent Poisson random variables is still a Poisson random variable, we know that $\sum_{k=1}^K b_k X_k(t)$ is  a Poisson random variable with mean $\mu_j$  as
\beq{env6666b}
    \sum_{k=1}^K b_k \rho_{k,j}=\sum_{k=1}^K b_k G_k \Lambda_k^\prime(1) r_{k,j}/F_j=\mu_j.
\eeq

As there are $F_j T$ class $j$ Poisson receivers, the probability that the $t$-th Poisson receiver is a class $j$ Poisson receiver is $F_j$. Thus, the expectation in the right-hand side of \req{env2233} can be computed by using the Poisson distribution, and we have
\bear{env7777}
    &&\sum_{k=1}^K b_k \ex[Y_k(t)] \nonumber\\
    &&\le \sum_{j=1}^J F_j \Big (\sum_{\tau=1}^{B-1} \tau e^{-\mu_j} \frac{{\mu_j}^\tau}{\tau!}+B (1-\sum_{\tau=1}^{B-1}e^{-\mu_j} \frac{{\mu_j}^\tau}{\tau!}) \Big ). \nonumber\\
\eear

For a system of CPRs to be stable (the departure rate must be the same as the arrival rate), we must have $G_k = \ex[Y_k(t)].$ Using this  in \req{env7777} leads to the upper bound for the stable region in \req{env4444}.
\eproof

\bex{spatialTRSA}{(Spatially coupled IRSA)}
    For the spatially coupled IRSA with $D$-multipacket reception capability in \cite{stefanovic2017asymptotic}, every time slot is a $D$-fold ALOHA and thus has the affine capacity envelope $(1,1,\ldots, 1)$ with the bound $D$. Since there is only one class of users $(K=1)$ and one class of receivers $(J=1)$ in \cite{stefanovic2017asymptotic}, the upper bound in \req{env4444} can be further simplified as follows:
    \bear{env4444IRSA}
        G &\le& \sum_{\tau=0}^{D-1} \tau e^{-G \Lambda^\prime(1) } \frac{{G \Lambda^\prime(1)}^\tau}{\tau!} \nonumber\\
        &&\quad+D \Big(1-\sum_{\tau=0}^{D-1}e^{-G \Lambda^\prime(1)} \frac{{G \Lambda^\prime(1)}^\tau}{\tau!} \Big ) \nonumber\\
        &=&D-\sum_{\tau=0}^{D-1} (D-\tau) e^{-G \Lambda^\prime(1) } \frac{{G \Lambda^\prime(1)}^\tau}{\tau!}.
    \eear
    This recovers the result in Theorem 1 of \cite{stefanovic2017asymptotic}.

One straightforward extension of the upper bound in \req{env4444IRSA} is to consider a mixture of $D$-fold ALOHA receivers in \cite{liu2021aloha}. Specifically, with probability $\pi_D$ (that satisfies $\sum_{D=1}^{D_{\rm max}} \pi_D=1$ for some positive integer $D_{\rm max}$), the (induced) Poisson receiver is selected from a $D$-fold ALOHA. For such a Poisson receiver, the success probability function (cf. \req{tfold1111}) is
\beq{mixALOHA1111}
    P_{\rm suc}(\rhog)=\sum_{D=1}^{D_{\rm max}}{ \pi_D \sum_{t=0}^{D-1} \frac{e^{-\rho} \rho^{t}}{t!}}.
\eeq
Now the bound $B$ in \rthe{capacity} is a random variable with the probability mass function
$$\pr (B=D)=\pi_D,$$
for $D=1, \ldots , D_{\rm max}$. Following the same argument in the proof of \rthe{capacity}, one can show that  the upper bound for the stable offered load is
\beq{mixALOHA2222}
    G \le \sum_{D=1}^{D_{\rm max}} \pi_D \Big (D-\sum_{\tau=0}^{D-1} (D-\tau) e^{-G \Lambda^\prime(1) } \frac{{G \Lambda^\prime(1)}^\tau}{\tau!} \Big).
\eeq
\eex

\bex{nearfar}{(Near-far SIC decoding)}
    For the setting with the capacity of near-far SIC decoding, every time slot can be modeled by the $\phi$-ALOHA receiver with
    \beq{phi7724b}
        \phi(n_1,n_2)=\left \{\begin{array}{cc}
            (n_1,n_2) & \mbox{if}\; (n_1,n_2) \le (1,1) \\
            (0,0) & \mbox{otheriwse}
        \end{array} \right ..
    \eeq
    The near-far SIC decoding is commonly used for modeling power domain NOMA \cite{shao2019noma,huang2021iterative}. There are three affine capacity envelopes for the  $\phi$-ALOHA receiver: (i) the vector $(1,0)$ with the bound 1, and (ii) the vector $(0,1)$ with the bound 1, and (iii) the vector $(1,1)$ with the bound 2. Suppose that there is only one class of receivers $(J=1)$. Then the first affine capacity envelope leads to the bound
    \beq{near1111}
        G_1\le 1-e^{-G_1 \Lambda^\prime_1(1) } ,
    \eeq
    and the second affine capacity envelope leads to the bound
    \beq{near2222}
    G_2\le 1-e^{-G_2 \Lambda^\prime_2(1) } ,
    \eeq
    and the third affine capacity envelope leads to the bound
    \bear{near3333}
        &&G_1+G_2 \nonumber\\
        &&\le 1 \cdot e^{-(G_1 \Lambda^\prime_1(1) +G_2 \Lambda^\prime_2(1)) } \frac{(G_1 \Lambda^\prime_1(1) +G_2 \Lambda^\prime_2(1))}{1!} \nonumber\\
        &&\quad +2\Big (1-e^{-(G_1 \Lambda^\prime_1(1) +G_2 \Lambda^\prime_2(1)) }(1\nonumber\\
        &&\quad\quad+\frac{(G_1 \Lambda^\prime_1(1) +G_2 \Lambda^\prime_2(1))}{1!}) \Big ).
    \eear
\eex

\bex{spatialTRSA_two_class}{(IRSA system with two classes of users and two classes of receivers)}
     We consider a system of coded Poisson receivers with two classes of users ($K=2$), two classes of receivers ($J=2$), and the success probability function $P_{{\rm suc}}(\rho)=e^{-\rho}$. This system is referred to as the {\em IRSA system with two classes of users and two classes of receivers} in \cite{chang2022stability}. As in \cite{chang2022stability}, we set $F_1=F_2=0.5$. We consider the following two packet routing policies (that correspond to two network slicing policies for resource allocation in uplink grant-free transmissions):
    \begin{description}
        \item[1)] Complete sharing: every packet has an equal probability to be routed to the two classes of receivers, i.e., $r_{11}=r_{22}=r_{12}=r_{21}=0.5$.
        \item[2)] Receiver reservation: class 1 packets are routed to the two classes of receivers with an equal probability, i.e., $r_{11}=r_{12}=0.5$, and class 2 packets are routed to the class 2 receivers, i.e., $r_{21}=0$, $r_{22}=1$.
    \end{description}
    Such a system is a CPR having three capacity envelopes: (i) the vector $(1,0)$ with the bound 1, and (ii) the vector $(0,1)$ with the bound 1, and (iii) the vector $(1,1)$ with the bound 1.

For the complete sharing policy and the affine capacity envelope $(1,1)$ with the bound 1, we have from \req{env6666} that
    \beq{IRSA2class1111}
        \mu_1=\mu_2=G_1\Lambda_1^\prime(1)+G_2\Lambda^\prime_2(1).
    \eeq
    Hence, by \req{env4444}, we have the outer bound
    \beq{IRSA2class2222}
        G_1+G_2\leq1-e^{-G_1\Lambda_1^\prime(1)-G_2\Lambda^\prime_2(1)}.
    \eeq

Similarly, for the receiver reservation policy and the affine capacity envelope $(1,1)$ with the bound 1, we have from \req{env6666} that
    \bear{IRSA2class1111r}
        \mu_1&=&G_1\Lambda_1^\prime(1), \nonumber\\
        \mu_2&=&G_1\Lambda_1^\prime(1)+2G_2\Lambda^\prime_2(1).
    \eear
Hence, by \req{env4444}, we have the outer bound
        \bear{IRSA2clas5555}
        G_1+G_2\leq1-\frac{1}{2}e^{-G_1\Lambda_1^\prime(1)}-\frac{1}{2}e^{-G_1\Lambda_1^\prime(1)-2G_2\Lambda_2^\prime(1)}.
    \eear

    On the other hand, for the receiver reservation policy and
    the affine capacity envelope $(0,1)$ with the bound 1, we have from \req{env6666} and \req{env4444} that
    $\mu_1=0$, $\mu_2=2G_2\Lambda_2^\prime(1)$  and
    \beq{IRSA2class9999}
        G_2\leq\frac{1}{2}-\frac{1}{2} e^{-2G_2\Lambda^\prime_2(1)}.
    \eeq

    We present the visualization of the bounds \req{IRSA2class2222}, \req{IRSA2clas5555}, and \req{IRSA2class9999} in this example in \rsubsec{irsatwoclass}.
\eex

\bsec{Convolutional coded Poisson receivers}{CCPR}

Inspired by the great performance of convolutional LDPC codes (see, e.g., \cite{kudekar2011threshold,yedla2012simple,mitchell2015spatially}) and the spatially coupled IRSA \cite{Liva2012spatially,stefanovic2017asymptotic}, in this section, we propose using the construction method for convolutional LDPC codes and spatially coupled IRSA to construct convolutional CPRs.

\bsubsec{Circular convolutional coded Poisson receivers}{CCPRc}

Recall that the ensemble of bipartite graphs in the $(T,G, \Lambda(x), R, F)$-CPR is constructed with the following parameters:
\begin{description}
    \item[1)] The total number of Poisson receivers $T$.
    \item[2)] The offered load vector $G=(G_1,G_2, \ldots, G_K)$.
    \item[3)] The degree distribution vector $\Lambda(x)=(\Lambda_1(x),\Lambda_2(x), \ldots, \Lambda_K(x))$.
    \item[4)] The $K \times J$ routing matrix $R=(r_{k,j})$.
    \item[5)] The partition vector $F=(F_1, F_2, \ldots, F_J)$.
\end{description}

Following the construction of convolutional LDPC codes, we first take $L$ independent copies of $(T,G^{(\ell)}, \Lambda(x), R, F)$-CPR with different offered load vectors, $G^{(\ell)}=(G^{(\ell)}_1, G^{(\ell)}_2, \ldots, G^{(\ell)}_K)$, $\ell=1,2, \ldots, L$, and concatenate the $L$ bipartite graphs to form a single bipartite graph. Then we rewire the edges in the concatenated bipartite graph. The receiver end of each edge in the $\ell^{th}$ copy is rewired to the corresponding receiver node in the $\hat \ell^{th}$ copy, where $\hat \ell$ is chosen uniformly in $[\ell, \ell\oplus (w-1)]$, where $1 \le w \le L$ is known as the ``smoothing'' window size in \cite{kudekar2011threshold}, and the $\oplus$ operator is the usual addition in a circular manner, i.e., for $1 \le \ell \le L$,
\beq{circdef}
    \ell\oplus (w-1)=\left \{\begin{array}{cc}
        \ell+ (w-1) & \mbox{if}\; \ell+(w-1) \le L \\
        \ell+(w-1)-L & \mbox{if}\; \ell+(w-1) > L
    \end{array} \right ..
\eeq
Also, we define the $\ominus$ operator as the usual subtraction in a circular manner, i.e.,
\beq{circdef2}
    \ell\ominus (w-1)=\left \{\begin{array}{cc}
        \ell-(w-1) & \mbox{if}\; \ell-(w-1) \ge 1 \\
        \ell-(w-1)+L & \mbox{if}\; \ell-(w-1) <1
    \end{array} \right ..
\eeq

We call such a system the circular convolutional CPR with $L$ stages. By viewing the class $k$ user nodes (resp. class $j$ receiver nodes) at the $\ell$ stage as the class $(k,\ell)$ users (resp. class $(j,\ell)$ receivers), the circular convolutional CPR with $L$ stages is a CPR with $KL$ classes of users and $JL$ classes of receivers. Moreover, the routing probability from a $(k,\ell)$ user node to a $(j,\hat \ell)$ receiver node is $r_{k,j}/w$ if $\hat \ell \in [\ell,\ell\oplus (w-1)]$ and 0 otherwise. Thus, the density evolution analysis in \rthe{mainext} can still be applied and we have the following corollary for the circular convolutional CPR with $L$ stages.

\bcor{mainextc}
    Consider the circular convolutional CPR with $L$ stages described in this section. Let $q_{k,\ell}^{(i)}$ be the probability that the {\em user end} of a randomly selected class $k$ edge in the $\ell^{th}$ stage has not been successfully received after the $i^{th}$ SIC iteration. Also, let
    \beq{mean4444mulrc}
        \rho_{k,j,\ell}^{(i)}= \sum_{\hat \ell=\ell \ominus (w-1)}^\ell q_{k,\hat \ell}^{(i)} G_{k}^{(\hat \ell)} \Lambda_{k}^\prime(1) \frac{1}{w} r_{k,j}/F_j,
    \eeq
    be the offered load of class $k$  users to a class $j$ receiver in the $\ell^{th}$ stage after the $i^{th}$ SIC iteration, and
    \beq{rho0000r}
        \trhojl^{(i)}=(\rho_{1,j,\ell}^{(i)}, \rho_{2,j,\ell}^{(i)}, \ldots, \rho_{K,j,\ell}^{(i)}),
    \eeq
    be the offered load vector to a class $j$ receiver in the $\ell^{th}$ stage after the $i^{th}$ SIC iteration.

    As $T \to \infty$, the success probability for a class $k$ user in the $\ell^{th}$ stage  after the $i^{th}$ SIC iteration converges to
    \bear{mean8888thumulamc}
        &&\tilde P_{{\rm suc},k,\ell}^{(i)}(G^{(1)}, G^{(2)}, \ldots, G^{(L)}) =1-\Lambda_k \Big  (1- \nonumber\\
        &&\sum_{\hat \ell=\ell}^{\ell\oplus (w-1)}\sum_{j=1}^J \frac{1}{w} r_{k,j}P_{{\rm suc},k,j}(\trhojhl^{(i)})\Big),
    \eear
    $k=1,2, \ldots, K$, $\ell=1,2, \ldots, L$, and $q^{(i)}_\ell=(q_{1,\ell}^{(i)}, q_{2,\ell}^{(i)}, \ldots, q_{K,\ell}^{(i)})$ can be computed recursively from the following equation:
    \beq{tag6666dmulmc}
        q_{k,\ell}^{(i+1)}=\lambda_{k} \Big (1- \sum_{\hat \ell=\ell}^{\ell\oplus (w-1)}\sum_{j=1}^J \frac{1}{w} r_{k,j}P_{{\rm suc},k,j}(\trhojhl^{(i)})) \Big),
    \eeq
    with $q^{(0)}_\ell=(1, 1, \ldots, 1)$. Also, under (A4), we can let $p_{k,\ell}^{(i+1)}=\lambda_k^{-1}(q_{k,\ell}^{(i+1)})$ and $p^{(i)}_\ell=(p_{1,\ell}^{(i)}, p_{2,\ell}^{(i)}, \ldots, p_{K,\ell}^{(i)})$. Then we can rewrite \req{tag6666dmulmc} as follows:
    \bear{tag6666bmulr2}
        p_{k,\ell}^{(i+1)}
        =1-\sum_{\hat \ell=\ell}^{\ell\oplus (w-1)}\sum_{j=1}^J \frac{1}{w} r_{k,j}P_{{\rm suc},k,j}(\trhojhl^{(i)}),\nonumber\\
    \eear
    with $p^{(0)}_\ell=(1, 1, \ldots, 1)$.
\ecor

One may refer to Appendix \ref{proof_of_mainextc} in the supplemental material for a detailed derivation. Similarly, the stability results in \rsec{stability} can also be applied.

\bsubsec{Stability and threshold saturation}{ts}

Now we construct the convolutional CPR from the circular convolutional CPR by setting the offered load vectors in the last $w-1$ stages to be the zero vector ${\bf 0}$. This is known as ``puncturing'' for convolutional LDPC codes.

\bdefin{CCPR}
    The convolutional $(T,G, \Lambda(x), R, F,w)$-CPR with $L$ stages is the circular convolutional CPR with $L$ stages when  $G^{(\ell)}=G$ for $\ell=1, 2,\ldots, L-w+1$, and $G^{(\ell)}={\bf 0}$ for $\ell=L-w+2, \ldots, L$.
\edefin

As the circular convolutional CPR with $L$ stages is  a CPR with $KL$ classes of users and $JL$ classes of receivers, the convolutional $(T,G, \Lambda(x), R, F,w)$-CPR with $L$ stages is a CPR with $K$ classes of users and $JL$ classes of receivers (by grouping the $(k,\ell)$ user nodes at the $L$ stages into a single class of user nodes, i.e., class $k$ user nodes). The notion of stability in \rsec{stability} can be defined the same way for the convolutional $(T,G, \Lambda(x), R, F,w)$-CPR with $L$ stages, i.e., it is stable if the probability that a packet is successfully received approaches 1 when the number of iterations goes to infinity.

Note that the normalized offered load for the convolutional $(T,G, \Lambda(x), R, F,w)$-CPR with $L$ stages is
\beq{special0000}
    \frac{\sum_{\ell=1}^L G^{(\ell)}T}{LT}=\frac{L-w+1}{L}G.
\eeq
Thus, as $L \to \infty$, the normalized offered load approaches $G$.

One of the most interesting phenomena for convolutional LDPC codes is {\em threshold saturation} (see, e.g., \cite{kudekar2011threshold, Liva2012spatially, mitchell2015spatially,xu2021some}). In the following theorem, we show an analogous result for the convolutional coded Poisson receivers.

\bthe{saturation}(Threshold saturation)
    Let $S$ be the stability region of the $(T,G, \Lambda(x), R, F)$-CPR and $S_L$ be the stability region of the convolutional $(T,G, \Lambda(x), R, F,w)$-CPR with $L$ stages. Then for any positive integer $L$,
    \beq{satu1111}
        S \subset S_{L} \subset S_{L-1}.
    \eeq
\ethe

In view of \rthe{saturation}, the stability region $S_L$ of the convolutional $(T,G, \Lambda(x), R, F,w)$-CPR with $L$ stages ``saturates'' when $L \to \infty$.

\bproof
Recall that the convolutional $(T,G, \Lambda(x), R, F,w)$-CPR with $L$ stages is the circular convolutional Poisson receiver with  $L$ stages when the last $w-1$ stages are ``punctured,'' i.e., $G^{(\ell)}=G$ for $\ell=1, 2,\ldots, L-w+1$, and $G^{(\ell)}={\bf 0}$ for $\ell=L-w+2, \ldots, L$. If the last $w$ stages are ``punctured,'' i.e., $G^{(\ell)}=G$ for $\ell=1, 2,\ldots, L-w$, and $G^{(\ell)}={\bf 0}$ for $\ell=L-w+1, \ldots, L$, then no packets are sent to the last stage and it can be removed. Thus, it reduces to the convolutional $(T,G, \Lambda(x), R, F,w)$-CPR with $L-1$ stages. As the success probability functions are decreasing in the offered load in (A1), the success probability in the convolutional $(T,G, \Lambda(x), R, F,w)$-CPR with $L$ stages is not larger than that of the convolutional $(T,G, \Lambda(x), R, F,w)$-CPR with $L-1$ stages. This shows that $S_{L} \subset S_{L-1}$.

Now consider the circular convolutional Poisson receiver with  $L$ stages when no  stages are ``punctured,'' i.e., $G^{(\ell)}=G$ for $\ell=1, 2,\ldots, L$. From the monotonicity, it is clear that the success probability of this system is not larger than that of the convolutional $(T,G, \Lambda(x), R, F,w)$-CPR with $L$ stages. We will argue that the success probability of this system is exactly the same as that of the $(T,G, \Lambda(x), R, F)$-CPR. As such, we have $S \subset S_{L}$.

When $G^{(\ell)}=G$ for $\ell=1, 2,\ldots, L$, $\rho_{k,j,\ell}^{(i)}$ in \req{mean4444mulrc} can be simplified as follows:
\beq{mean4444mulrcs}
    \rho_{k,j,\ell}^{(i)}= {G_k}\Lambda_{k}^\prime(1) r_{k,j}/F_j \sum_{\hat \ell=\ell \ominus (w-1)}^\ell q_{k,\hat \ell}^{(i)}.
\eeq
We now use induction to show that $q_{k,\ell}^{(i)}=q_k^{(i)}$ for all $i$. Since $q_{k,\ell}^{(0)}=1$ and $q_k^{(0)}=1$, we have from \req{mean4444mulrcs} that
\beq{mean4444mulrcs1}
    \rho_{k,j,\ell}^{(1)}= {G_k}\Lambda_{k}^\prime(1) r_{k,j}/F_j.
\eeq

In view of \req{mean6666dmulr} and \req{rho0000r}, we have
$$\trhojl^{(1)}=q^{(1)}_{\ell} \circ \trhoj.$$
It is easy to see from \req{tag6666dmulmc} and \req{tag6666dmulm} that $q_{k,\ell}^{(1)}=q_k^{(1)}$ for all $\ell=1,2, \ldots, L$. By inducting  on the number of iterations $i$, we then have $q_{k,\ell}^{(i)}=q_k^{(i)}$ for all $i$. Because of \req{mean8888thumulamc} and \req{mean8888thumulam}, we conclude that the success probability of the circular convolutional Poisson receiver without puncturing is the same as that of the $(T,G, \Lambda(x), R, F)$-CPR.
\eproof

In the case where there is only one class of users and one class of receivers (the single-class system), the stability region of the convolutional $(T,G, \Lambda(x), R, F,w)$-CPR with $L$ stages can be reduced to a set in $\mathbb{R}$. The supremum of this set represents the percolation threshold of the CCPR. A direct finding from \rthe{saturation} is the monotonic decrease of the percolation threshold as the number of stages $L$ increases.

\bex{irsacon}{(Convolutional IRSA)}
    For slotted ALOHA, the success probability function $P_{{\rm suc}}(\rho)=e^{-\rho}$. Suppose that the degree distribution $\Lambda(x)=x^d$, i.e., each packet is transmitted exactly $d$ times. For such a degree distribution, we have $\lambda(x)=x^{d-1}$. Assume $w=2$, $K=1$, and $J=1$. It follows from \req{tag6666dmulmc} in \rcor{mainextc} that for $\ell=2,\ldots, L-2$,
    \bear{tag6666dmulmsb1ll}
        &&q_{1}^{(i+1)}\nonumber\\
        &&=\Big (1-\frac{1}{2}e^{-\frac{1}{2}G d q_{1}^{(i)}} -\frac{1}{2}e^{-\frac{1}{2}G d ({q_{1}^{(i)}+q_{2}^{(i)}})}\Big )^{d-1},\nonumber\\
    \eear
    \bear{tag6666dmulms1ll}
        &&q_{\ell}^{(i+1)}\nonumber\\
        &&=\Big (1-\frac{1}{2}e^{-\frac{1}{2}G d ({q_{\ell-1}^{(i)}+q_\ell^{(i)}})} -\frac{1}{2}e^{-\frac{1}{2}G d ({q_{\ell}^{(i)}+q_{\ell+1}^{(i)}})}\Big )^{d-1},\nonumber\\
    \eear
    and
    \bear{tag6666dmulmsc1ll}
        &&q_{L-1}^{(i+1)}\nonumber\\
        &&=\Big (1-\frac{1}{2}e^{-\frac{1}{2}G d ({q_{L-2}^{(i)}+q_{L-1}^{(i)}})}-\frac{1}{2}e^{-\frac{1}{2}G d q_{L-1}^{(i)}} \Big )^{d-1},\nonumber\\
    \eear
    with the initial condition $q_k^{(0)}=1$, $k=1,2,\ldots, L-1$. It can be shown that from Example 5 of \cite{chang2022stability} that the percolation thresholds are $G=0.9179, 0.9767, 0.9924, 0.9973$ for $d=3,4,5,6$. We evaluate the thresholds for the system described in this example with $w=2,3,4$ in \rsec{irsaoneclass}.
\eex

\bsec{Percolation thresholds of systems with a single class of users}{saturation}

In this section, we focus on CPR systems with only one class of users and one class of receivers. We derive the potential function of such a system of CPR and the associated percolation thresholds of both CPRs and CCPRs.

\bsubsec{The single-system threshold}{SAsystem}

Since there is only one class of users and one class of receivers, \req{tag6666dmulm} and \req{key1111} reduce to the following form:
\bear{tag6677}
    p^{(i)}&=& 1-P_{\rm suc}(q^{(i)} G \Lambda^\prime(1)), \label{eq:tag6677a}\\
    q^{(i+1)}&=&\lambda(p^{(i)}) \label{eq:tag6677b}.
\eear

We combine the two equations in
\req{tag6677a} and \req{tag6677b} to form the following recursive equation:
\beq{key4444}
    p^{(i+1)}=1-P_\text{suc}(\lambda(p^{(i)})G\Lambda^\prime(1)),
\eeq
with $p^{(0)}=1$. For every packet to be decoded successfully, we need to ensure
that
\beq{key4456}
    \lim_{i \to \infty} p^{(i)}=0.
\eeq
We are interested in finding out the maximum offered load $G$ such that \req{key4456} is satisfied. Let
\bear{fh}
    f(p;G)&=&1-P_\text{suc}(pG\Lambda^\prime(1)), \label{eq:fhbbbb}\\
    h(p)&=&\lambda(p). \label{eq:fhaaaa}
\eear

Then \req{key4444} can be written as follows:
\beq{scalaradmissible}
    p^{(i+1)}=f(h(p^{(i)});G).
\eeq

In \cite{yedla2012simple}, the recursion of the form in \req{scalaradmissible} is said to be a {\em scalar admissible system} characterized by a pair of functions  $(f, h)$ that satisfy the four properties in \rdef{scalar_admissible_system} below.

\bdefin{scalar_admissible_system}(cf. Def. 1 in \cite{yedla2012simple})
    The scalar admissible system $(f,h)$ parameterized by $G \geq 0$ is defined by the recursion $p^{(i+1)}=f(h(p^{(i)});G)$, where $f$ and $h$ satisfy the following four properties:
    \begin{description}
        \item[(P1)] $f:[0,1]\times[0,\infty)\to[0,1]$ is strictly increasing in both $p$ and $G$.
        \item[(P2)] $h:[0,1]\to[0,1]$ satisfies $h^\prime(p)>0$ for $p\in(0,1]$.
        \item[(P3)] $f(0;G)=f(p;0)=h(0)=0$.
        \item[(P4)] $f$ has continuous second derivatives on $[0,1]\times[0,\infty)$ w.r.t. all arguments, so is $h$ on $[0,1]$.
    \end{description}
\edefin

In the following lemma, we show that the $(T, G, \Lambda(x), R, F)$-CPR with one class of users and one class of receivers is indeed a scalar admissible system under the assumptions in (A1)-(A4).

\blem{CPR_SA}
    Under the assumptions in (A1) and (A4), a system of CPR described by the density evolution equation in \req{key4444} is the scalar admissible system $(f,h)$ with $f$ and $h$ specified in \req{fhbbbb} and \req{fhaaaa}, respectively.
\elem

\bproof
It suffices to show that $f$ in \req{fhbbbb} and $h$ in \req{fhaaaa} satisfy (P1)-(P4) of \rdef{scalar_admissible_system}.

(P1) Since $P_\text{suc}$ is (strictly) decreasing in (A1), we have for $p\in[0,1]$ and $G \in [0, \infty)$,
\bearn
    \frac{\partial f(p;G)}{\partial p}&=&-\frac{\partial P_\text{suc}(pG\Lambda^\prime(1))}{\partial p}\cdot G\Lambda^\prime(1)>0,\\
    \frac{\partial f(p;G)}{\partial G}&=&-\frac{\partial P_\text{suc}(pG\Lambda^\prime(1))}{\partial G}\cdot p\Lambda^\prime(1)>0.
\eearn

(P2) Since every packet is transmitted at least twice in (A4), we have $\Lambda^{\prime\prime}(p)>0$, $\forall p\in [0,1]$. Also, since $\Lambda'(1)$ is positive, $h'(p)=\lambda^\prime(p)=\Lambda^{\prime\prime}(p)/\Lambda^{\prime}(1)>0$.

(P3) Since $P_\text{suc}(0)=1$, and $\lambda$ is a polynomial without constant terms,
\bearn
    f(0,G)&=&1-P_\text{suc}(0)=0,\\
    f(p,0)&=&1-P_\text{suc}(0)=0,\\
    h(0)&=&\lambda(0)=0.
\eearn

(P4) is obvious from (A1).
\eproof

One of the most powerful tools for analyzing the stability of the scalar admissible system $(f,h)$ is the {\em potential function} in \cite{yedla2012simple}.

\bdefin{potential_function}
    The function $U(p;G)$ is called the {\em potential function} of the scalar admissible system $(f,h)$ in \cite{yedla2012simple} if it satisfies
    \bear{potential1111}
        p^{(i+1)}-p^{(i)}&=&f(h(p^{(i)});G)-p^{(i)}\nonumber\\
        &=&-\frac{1}{h^\prime(p^{(i)})}\frac{\partial U(p;G)}{\partial p}\bigg|_{p=p^{(i)}}.
    \eear
\edefin

To ease the representation, for a function $U(p;G)$ with two variables, we use $U^\prime(p;G)$ to denote the partial differentiation with respect to the {\em first} variable, given by $$U^\prime(p;G)=\frac{\partial U(p;G)}{\partial p}.$$

Rewrite \req{potential1111} as follows:
\beq{potential2222}
    U^\prime (p;G)\bigg|_{p=p^{(i)}}=-{h^\prime(p^{(i)})} (f(h(p^{(i)});G)-p^{(i)}).
\eeq
Integrating both sides of \req{potential2222} yields
\bear{potential_int_scalar}
    U(p;G)&=&\int_0^p h'(z)(z-f(h(z);G))dz\nonumber\\
    &=&ph(p)-H(p)-F(h(p);G),
\eear
where
\beq{potential3333}
    F(p;G)=\int_0^p f(z;G)dz,
\eeq
and
\beq{potential4444}
    H(p)=\int_0^p h(z)dz.
\eeq
To see how the potential function can be used for analyzing the stability of a scalar admissible system, let us consider the {\em single-system threshold} \cite{yedla2012simple} defined  below:
\beq{Gsstar_def}
    G_s^*=\sup\{G\in[0,1]|U'(p;G)>0, \forall p\in(0,1]\}.
\eeq
Then for all $G \le G_s^*$, the system is stable. To see this, note from (P1) of \rdef{scalar_admissible_system} that $U'(p;G)>0$ for all $p \in (0,1]$ and $G \le G_s^*$. Also, note from (P3) of \rdef{scalar_admissible_system}, that is, $f(0;G)=f(p;0)=h(0)=0$,  that $p=0$ is a solution of $p=f(h(p);G)$. Since  $h^\prime(p)>0$ for $p\in(0,1]$ ((P2) in \rdef{scalar_admissible_system}), one can see from \req{potential1111} that $p^{(i+1)}<p^{(i)}$ for all $i$. Thus, $p^{(i)}$ (with $p^{(0)}=1$) converges to $0$ as $i \to \infty$ for all $G\le G_s^*$.

Since we have shown in \rlem{CPR_SA} that the $(T, G, \Lambda(x), R, F)$-CPR with one class of users and one class of receivers is a scalar admissible system with $f$ and $h$ specified in \req{fhbbbb} and \req{fhaaaa}, we can use \req{potential_int_scalar} to derive the potential function $U(p;G)$ for $(T, G, \Lambda(x), R, F)$-CPRs, and then use that to show the following stability result.

\bthe{CCPR_potential}
Under the assumptions in (A1)-(A4), the potential function of the $(T, G, \Lambda(x), R, F)$-CPR with one class of users and one class of receivers is given by
\bear{CCPR_potential}
    U(p;G)=&&\lambda(p)(p-1)-\frac{\Lambda(p)}{\Lambda^\prime(1)}\nonumber\\
           &&+\frac{1}{G\Lambda^\prime(1)}\int_0^{G\Lambda^\prime(1)\lambda(p)}P_\text{suc}(\rho)d\rho.
\eear

Moreover, if the inverse function of $P_\text{suc}(\cdot)$ exists and is continuous and decreasing on $[0,1]$, then for all $G \le G_s^*$, the system is stable, where $G_s^*$ in \req{Gsstar_def} has the following representation:
\bear{Gsstar}
    G_s^*=\inf_{p\in[0,1]}\frac{P^{-1}_\text{suc}(1-p)}{\Lambda^\prime (p)}.
\eear
\ethe

\bproof
In view of  \req{fh} and\req{potential3333}, integrating the function $f$ w.r.t. $z$ from 0 to $p$ yields
\bear{potentialFFFF}
    F(p;G)&&=\int_0^p 1-P_\text{suc}(zG\Lambda^\prime(1))dz\nonumber\\
    &&=p-\frac{1}{G\Lambda^\prime(1)}\int_0^{G\Lambda^\prime(1)p}P_\text{suc}(\rho)d\rho.
\eear

Also, note that $h(p)=\lambda(p)=\Lambda^\prime(p)/\Lambda^\prime(1)$, $H(p)=\Lambda(p)/\Lambda^\prime(1)$. Using \req{potential_int_scalar}, we obtain \req{CCPR_potential}.

From \req{potential1111}, the partial derivative of the potential function w.r.t. its first variable, known as the {\em balance function} has the following form:
\beq{CCPR_potential_prime}
    U'(p;G)=\lambda^\prime (p)(p-1+P_\text{suc}(G\Lambda^\prime(p)),
\eeq
where $\Lambda^\prime(p)=\Lambda^\prime(1)\lambda(p)$. Since we assume that $P_\text{suc}^{-1}$ exists, from \req{Gsstar_def} and \req{CCPR_potential_prime}, we have that
\bear{CPRpotential0000}
    G_s^*&=&\sup\{G\in[0,\infty)|U'(p;G)>0, \forall p\in(0,1]\}\nonumber\\
    &=&\sup\{G\in[0,\infty)|P_\text{suc}(G\Lambda^\prime (p))>1-p,\nonumber\\
    &&\forall p\in(0,1]\}\nonumber\\
    &=&\inf_{p\in[0,1]}\frac{P^{-1}_\text{suc}(1-p)}{\Lambda^\prime (p)}.
\eear
\eproof

The following example shows the potential functions of the $D$-fold ALOHA system.

\bex{DfoldALOHA_potential}{($D$-fold ALOHA)}
For the $D$-fold ALOHA system described by \req{phi7722} and \req{tfold1111} with $\Lambda(x)=x^d$, we use \req{CCPR_potential} and integration by parts (as in Theorem 3 of \cite{Manuel2023irregular}) to evaluate the potential function as follows:
\bear{udaloha}
    &&U(p;G)=\frac{1}{d}\Big ((d-1)p^d-dp^{d-1}+\frac{1}{G}\bigg(D-\nonumber\\
    &&\quad e^{-Gdp^{d-1}}\sum_{\tau=0}^{D-1}(D-\tau)\frac{(Gdp^{d-1})^\tau}{\tau!}\bigg) \Big)..\nonumber\\
\eear
\eex

\bsubsec{The saturation theorem}{mainresult}

In this section, we study the stability of the convolutional $(T, G, \Lambda(x), R, F,w)$-CPR with $L$ stages. The density evolution equation for the convolutional $(T,G, \Lambda(x), R, F,w)$-CPR is given by \req{tag6666bmulr2} in \rcor{mainextc}. For $K=J=1$, by \req{mean3344mulr}, \req{mean4444mulrc}, and the relation  $q_{\ell}^{(i)}=\lambda(p_\ell^{(i)})$, \req{tag6666bmulr2} can be simplified as
\beq{DEofCCPRoneclass}
    p_{\ell}^{(i+1)}=1-\sum_{\hat{\ell}=\ell}^{\ell\oplus(w-1)}\frac{1}{w}P_{\text{suc}}\bigg(\frac{1}{w}\sum_{\tilde \ell=\hat{\ell}\ominus(w-1)}^{\hat{\ell}} \lambda(p_{\tilde \ell}^{(i)})G^{(\tilde \ell)}\Lambda^\prime(1)\bigg),
\eeq
for $\ell=1,\ldots,L$. Moreover, since $G^{(\tilde \ell)}=G$ for $\tilde \ell=1,2, \ldots, L-w+1$ and $G^{(\tilde \ell)}=0$ for $\tilde \ell=L-w+2, \ldots, L$,
\req{DEofCCPRoneclass} can be further simplified as follows:

\beq{simplified}
    p_{\ell}^{(i+1)}=1-\sum_{\hat{\ell}=\ell}^{\ell+(w-1)}\frac{1}{w}P_{\text{suc}}\bigg(\frac{G\Lambda^\prime(1)}{w}\sum_{\tilde \ell=\max[1,\hat{\ell}-(w-1)]}^{\min[L-w+1,\hat{\ell}]}\lambda(p_{\tilde \ell}^{(i)})\bigg),
\eeq
for $\ell=1,2, \ldots, L-w+1$.

Denote by $G_{conv}^*(L,w)$ the percolation threshold of the convolutional $(T, G, \Lambda(x), R, F,w)$-CPR system with $L$ stages. The only way to know the exact value of $G_{conv}^*(L,w)$ is to evaluate the density evolution equations in \req{DEofCCPRoneclass}. In this section, we prove a lower bound of $G_{conv}^*(L,w)$ by using the potential function of the corresponding $(T, G, \Lambda(x), R, F)$-CPR (under certain conditions of the window size $w$). Such a stability result for the convolutional $(T, G, \Lambda(x), R, F,w)$-CPR system with $L$ stages is called the {\em saturation theorem} in this paper.

In addition to the single-system threshold $G_s^*$ in \req{Gsstar_def}, we define the {\em potential threshold} of the $(T, G, \Lambda(x), R, F)$-CPR. We will show that the potential threshold is a lower bound of $G_{conv}^*(L,w)$ in the saturation theorem.

\bdefin{potential_threshold}(cf. Def. 6 in \cite{yedla2012simple})
    Consider the scalar admissible system described in \rdef{scalar_admissible_system} with the potential function $U(p;G)$ in
     \req{potential_int_scalar}.  The potential threshold of the scalar admissible system, denoted by $G_{conv}^*$, is defined below:
    \beq{potential_threshold_def}
        G_{conv}^*=\sup\{G\in[0,1]|\min_{p\in[0,1]}U(p;G) \geq 0\}.
    \eeq
\edefin

To give the specific conditions for the window size of the saturation theorem, we define the following terms.

\bdefin{minimum_unstable}(cf. Def. 5, and 6 in \cite{yedla2012simple})
    Consider the scalar admissible system $(f,h)$ in \rdef{scalar_admissible_system} with the potential function $U(p;G)$ in
     \req{potential_int_scalar}.
    \begin{description}
        \item[(i)] For $G>G_s^*$, the minimum unstable fixed point is the number
        \beq{unstable1111}
        u(G) = \sup\{\tilde{p}\in[0,1]|U'(p;G)\geq 0, p\in[0,\tilde{p}]\}.
        \eeq
        \item[(ii)] The energy gap of the scalar admissible system for $G\in(G_s^*,1]$ is the number
        \beq{unstable2222}
        \Delta E(G)=\min_{p\in[u(G),1]}U(p;G).
        \eeq
        \item [(iii)] A constant $K_{f,h}=||h'||_\infty + ||h'||^2_\infty||f'||_\infty+||h''||_\infty$,
    where $$||h||_\infty=\sup_{x\in[0,1]} |h(x)|$$ for functions $h:[0,1]\to\mathbb{R}$.
    \end{description}
\edefin

Now we state the saturation theorem.

\bthe{saturation_singleclass} (The saturation theorem for the convolutional $(T, G, \Lambda(x), R, F,w)$-CPR)
    Consider the single-class ($K=J=1$) convolutional $(T, G, \Lambda(x), R, F,w)$-CPR system with $L$ stages governed by the density evolution recursion in \req{simplified}. If $G<G_{conv}^*$ and
    \beq{unstable3333}
        w>\frac{K_{f,h}}{\Delta E(G)},
    \eeq
    then the only fixed point of \req{simplified} is $\mathbf{p}=\mathbf{0}$. As such, the convolutional $(T, G, \Lambda(x), R, F,w)$-CPR system with $L$ stages is stable for all $G < G_{conv}^*$.
\ethe

Our proof is analogous to that for the stability of spatially coupled LDPC codes (Theorem 1 in \cite{yedla2012simple}), and it requires a sequence of lemmas and definitions of terms. Here we outline the basic steps of the proof. The detailed proof can be found in Appendix \ref{proof_of_mainextc} of the supplemental material.

\noindent (i) Let $\tilde{L}=L-w+1$. We reverse all the indices of the recursion relation \req{simplified}. Then, the recursion \req{simplified} becomes

\beq{reversed}
    p_\ell^{(i+1)}=1-\sum_{\hat{\ell}=\tilde{L}+1-\ell}^{\tilde{L}+w-\ell}\frac{1}{w}P_{\text{suc}}\bigg(\frac{G\Lambda^\prime(1)}{w}\sum_{\tilde \ell=\max[1,\tilde{L} - \hat{\ell} + 1]}^{\min[\tilde{L},\tilde{L} - \hat{\ell} + w]}\lambda(p_{\tilde \ell}^{(i)})\bigg).
\eeq

\noindent (ii) If $h$ is a function with one variable and $\mathbf{p}=(p_1,\ldots,p_{\tilde{L}})\in\mathbb{R}^{\tilde{L}}$ is a vector, we define
$$\mathbf{h}(\mathbf{p})=(h(p_1), h(p_2),\ldots,h(p_{\tilde{L}}))$$
and
$$\mathbf{f}(\mathbf{p};G)=(f(p_1;G), f(p_2;G),\ldots,f(p_{\tilde{L}};G))$$
for the ease of representation. Let $\mathbf{p}^{(i)}=(p^{(i)}_1,p^{(i)}_2, \ldots, p^{(i)}_{\tilde{L}})$, then \req{reversed} can be written as the following matrix form:
\beq{sc0000}
    \mathbf{p}^{(i+1)}=\mathbf{A}_2\mathbf{f}(\mathbf{A}_2^T\mathbf{h}(\mathbf{p}^{(i)});G),
\eeq

where $\mathbf{A}_2$ is an $\tilde{L}\times L$ matrix defined as:
\beq{scmatrixA}
    \mathbf{A}_2=\frac{1}{w}\begin{bmatrix}
        1      & 1      & \ldots  & 1      & 0      & 0      & \ldots  & 0      \\
        0      & 1      & 1      & \ldots  & 1      & 0      & \ldots  & 0      \\
        \vdots & \ddots & \ddots & \ddots & \ddots & \ddots & \ddots & \vdots \\
        0      & \ldots  & 0      & 1      & 1      & \ldots  & 1      & 0      \\
        0      & \ldots  & 0      & 0      & 1      & 1      & \ldots  & 1      \\
    \end{bmatrix},
\eeq

where all the rows of $\mathbf{A}_2$ contain $w$ ones, and the functions $f$ and $h$ are specified in \req{fhbbbb} and \req{fhaaaa}, respectively. Such a system of recursive equations is called the {\em basic spatially-coupled} system in \cite{yedla2012simple}, up to that the matrix $\mathbf{A}_2$ is transposed. Like the scalar admissible system, the basic spatially-coupled system is also characterized by the two functions $f$ and $h$.

\noindent (iii) Consider the recursive equations in the following matrix form:
\beq{onesided0000}
    \mathbf{p}^{(i+1)}=\mathbf{A}^T\mathbf{f}(\mathbf{A}\mathbf{h}(\mathbf{p}^{(i)});G),
\eeq
where $\mathbf{A}$ is an $L\times L$ matrix defined as:
\beq{onesidedmatrixA}
    \mathbf{A}=\frac{1}{w}\begin{bmatrix}
        1      & 1      & \ldots  & 1      & 0      & \ldots  & 0 \\
        0      & 1      & 1      & \ldots  & 1      & \ddots & \vdots \\
        \vdots & \ddots & \ddots & \ddots & \ddots & \ddots & 0 \\
        0      & \ldots  & 0      & 1      & 1      & \ldots  & 1 \\
        0      & 0      & \ldots  & 0      & 1      & \ldots  & 1 \\
        0      & 0      & \ldots  & 0      & 0      & 1      & 1 \\
        0      & 0      & \ldots  & 0      & 0      & 0      & 1
    \end{bmatrix},
\eeq
where the 1st, 2nd, \ldots, and the $(L-w+1)$-th rows of $\mathbf{A}$ contain $w$ ones, and the functions $f$ and $h$ are specified in \req{fhbbbb} and \req{fhaaaa}, respectively. Such a system of recursive equations is called the {\em vector one-sided spatially-coupled recursion} system in \cite{yedla2012simple} and \cite{Yedla2012vector}. We will briefly call this kind of system a {\em one-sided system} in this paper. Like the scalar admissible system, the one-sided system is also characterized by the two functions $f$ and $h$.

One may illustrate that the one-sided system gives a component-wise upper bound of the basic spatially-coupled system for $\ell=1,\ldots,\tilde{L}$. Hence, the percolation threshold of the basic spatially-coupled system is not smaller than that of the one-sided system. Once we prove the saturation theorem for the one-sided system, the results will also apply to the basic spatially-coupled system.

\noindent (iv) The {\em potential function of the one-sided system} has the following representation:
\beq{onesided1111}
    U(\mathbf{p};G)=\mathbf{h}(\mathbf{p})^T\mathbf{p}-H(\mathbf{p})-F(\mathbf{A}\mathbf{h}(\mathbf{p});G),
\eeq
where
\beq{onesided2222}
    H(\mathbf{p})=\sum_\ell H(p_\ell),
\eeq
and
\beq{onesided3333}
    F(\mathbf{p};G)=\sum_\ell F(p_\ell;G),
\eeq
with the functions $F$ and $H$ being defined in \req{potential3333} and \req{potential4444}, respectively. This shows how the potential function of the one-sided system is coupled with that of the scalar admissible system.

\noindent (v) Reproduce the properties of the potential function in Lemma 3 and Lemma 4 of \cite{yedla2012simple} to show a specific relation between the potential function of the one-sided system and that of the scalar admissible system. Specifically, for a non-decreasing vector $\mathbf{p}=(p_1,\ldots,p_{L})\in[0,1]^{L}$, define the shift operator $\mathbf{S}:\mathbb{R}^{L}\to\mathbb{R}^{L}$ so that $[\mathbf{S}\mathbf{p}]=(0,p_1,p_2,\ldots,p_{{L}-1})$. Then
\beq{onesided4444}
    U(\mathbf{S}\mathbf{p};G)-U(\mathbf{p};G)=-U(p_{L};G).
\eeq

\noindent (vi) Reproduce the bound for the norm of the Hessian $U''(\mathbf{p};G)$ in Lemma 5 of \cite{yedla2012simple}, i.e.,
\beq{onesided5555}
    ||U''(\mathbf{p};G)||_\infty\leq K_{f,h},
\eeq
where the constant $ K_{f,h}$ is specified in \rdef{minimum_unstable} (iii).

\noindent (vii) Suppose $\mathbf{p}\neq\mathbf{0}$ is a fixed point of \req{onesided0000}. As in the proof of Theorem 1 of \cite{yedla2012simple}, one can use the Taylor series (with the bound for the norm of the Hessian $U''(\mathbf{p};G)$) to show that the inner product of the gradient $U'(\mathbf{p};G)$ with a specific direction $\mathbf{S}\mathbf{p}-\mathbf{p}$ is smaller than 0 for $G < G_{conv}^*$ and $w$ satisfying the condition in \req{unstable3333}. Since each component of the vector $\mathbf{S}\mathbf{p}-\mathbf{p}$ is not greater than 0, there must exist a positive component, say the $\ell_0$-th component, of the gradient $U'(\mathbf{p};G)$. As such, one more iteration further reduces the value of $p_{\ell_0}$. This then leads to a contradiction and thus shows that $\mathbf{p}=\mathbf{0}$ is the only fixed point of \req{DEofCCPRoneclass}.

Note that if $L$ is large enough, one can always choose a window size $w$ large enough to satisfy \req{unstable3333}.
An immediate consequence of \rthe{saturation_singleclass} is that $G_{conv}^*(L,w)\geq G_{conv}^*$ if $w>K_{f,h}/\Delta E(G)$, as \rthe{saturation_singleclass} provides a sufficient condition for the stability of the convolutional $(T, G, \Lambda(x), R, F,w)$-CPR system with $L$ stages. However, it is also possible that $G_{conv}^*(L,w)$ is larger than $G_{conv}^*$ even when the condition in \req{unstable3333} is not satisfied. This is due to the puncturing effect that reduces the offered load. In addition, the condition for $w$ in \rthe{saturation_singleclass} often results in large window size. However, our numerical results in \rsec{irsaoneclass} demonstrate that $|G_{conv}^*(L,w)-G_{conv}^*|<0.0001$ can be achieved in some cases.

\bsubsec{The upper bound by the area theorem}{notable}

In this section, we consider another threshold $G_{up}^*$ that is defined as the unique positive solution of the equation $U(1;G)=0$. For the potential function of a system of CPR in \req{CCPR_potential}, $G_{up}^*$ is the unique positive solution to the following equation:
\beq{barg1111}
    G=\int_0^{G \Lambda^\prime(1)} P_\text{suc}(\rho)d\rho.
\eeq

Now we show by the area theorem \cite{ashikhmin2004extrinsic} that $G_{up}^*$ is an upper bound for the stable offered load of a system of CPR. For  $p^{(i)}$ and $q^{(i)}$ in the two recursive mappings in \req{tag6677a} and \req{tag6677b} to converge to 0 as $i \to \infty$, the area theorem in \cite{ashikhmin2004extrinsic} provides a necessary condition that requires the sum of the two areas under the two mappings to be not greater than 1, i.e.,
\beq{tag8888NOMA}
    \int_0^1 (1-P_{{\rm suc}}(q G \Lambda^\prime(1)) dq+ \int_0^1 \lambda(p) dp \le 1.
\eeq
Since $\lambda(p)=\Lambda^\prime (p)/\Lambda^\prime (1)$, $\int_0^1 \lambda(p) dp= 1/\Lambda^\prime (1)$. Let $\rho=q G \Lambda^\prime(1)$. We can write the inequality in \req{tag8888NOMA} as follows:
\beq{tag9999NOMA}
    G \le \int_0^{G \Lambda^\prime(1)}  P_{{\rm suc}}(\rho)d\rho.
\eeq

For the $D$-fold ALOHA system described by \req{phi7722} and \req{tfold1111}, $G_{up}^*$ is the solution to the following equation:
\bear{DfoldALOHAupperbound}
    G = D-\sum_{\tau=0}^{D-1} (D-\tau)e^{-Gd}\frac{(Gd)^\tau}{\tau!}.
\eear

This is the same as the upper bound in Theorem 1 of \cite{stefanovic2017asymptotic} for the spatially coupled $D$-fold ALOHA system. It is also the same as the upper bound in \rsec{upper} when the underlying channel is modeled by a $\phi$-ALOHA receiver \cite{liu2021aloha}. In view of the extended upper bound for a mixture of $D$-fold ALOHA in \rex{spatialTRSA}, we have the following corollary.

\bcor{upperbound}
    Suppose that the $T$ Poisson receivers of the  $(T, G, \Lambda(x), R, F)$-CPR with one class of users and one class of receivers in \req{CCPR_potential}  are induced from a mixture of $D$-fold ALOHA in \rex{spatialTRSA}. Then the stable offered load of such a system is bounded above by $G_{up}^*$, where $G_{up}^*$ is the solution to \req{barg1111}. As such, $G_{up}^*$ is an upper bound for the percolation threshold of the convolutional $(T, G, \Lambda(x), R, F,w)$-CPR  with an infinite number of stages, i.e.,
    $$\lim_{L \to \infty}G_{conv}^*(L,w)<G_{up}^*$$
    regardless of the choice of the degree distribution $\Lambda(x)$ and the window size $w$.
\ecor

In the following theorem, we show the single-system threshold $G_s^*$ in \req{Gsstar_def} is smaller than the potential threshold $G_{conv}^*$, and the potential threshold $G_{conv}^*$ is smaller than the upper bound $G_{up}^*$.

\bthe{upperboundbypotential}
    Consider the scalar admissible system $(f,h)$ in \rdef{scalar_admissible_system} with the potential function $U(p;G)$. Then
    \beq{main1234}
        G_s^*<G_{conv}^*<G_{up}^*.
    \eeq
\ethe

\bproof
(i) First, we demonstrate that $G_s^*<G_{conv}^*$. From the integral representation in \req{potential_int_scalar} and the fact that the function $f$ is strictly increasing in $G$ from (P1), we deduce that $U(0;G)=0$ for all $G$ and $U(p;G)$ is strictly decreasing in $G$. From the definition given in \req{Gsstar_def} of $G_s^*$, it is clear that $U'(p;G)>0$ when $G\leq G_s^*$. Therefore, the function $U(p;G)$ is strictly increasing in $p$ when $G\leq G_s^*$. This implies that $U(p;G) > 0$ for all $p\in(0,1)$ and  $G\leq G_s^*$. Since $U(p;G)$ decreases for $G$, when $U(p;G)$ becomes 0 for some value of $G$ and $p$ is not 0, such a $G$ must be greater than $G_s^*$. The supremum of all such possible $G$ values results in $G_s^*<G_{conv}^*$.

(ii) Now we show that $G_{conv}^*<G_{up}^*$.  By (A4), we have $\Lambda^\prime(1)>0$ and $\lambda^\prime(1)>0$. Since $P_\text{suc}(G\Lambda^\prime(1))>0$ from (A1), we have
\begin{equation*}
    U'(1;G)=\lambda^\prime(1)P_\text{suc}(G\Lambda^\prime(1))>0.
\end{equation*}

Thus, $p=1$ cannot be a local minimum of $U(p;G)$ for $G>0$. As such,
\beq{potential_ineq}
    U(1;G_{conv}^*)>\min_{p\in[0,1]}U(p;G_{conv}^*)=0.
\eeq

As argued in (i), $U(1;G)$ is strictly decreasing in $G$. Thus, we have from \req{potential_ineq} that $U(1;G)>0$ for all $G \le G_{conv}^*$. This shows that $G_{conv}^*<G_{up}^*$ if $G_{up}^*$ exists. It remains to show the existence of the unique positive solution $G_{up}^*$ of the equation $U(1;G)=0$. Since $U(1;G)$ is continuous and decreasing for $G>0$, and $f(p;G)$ is strictly increasing in both $p$ and $G$, $U(1;G)\to -\infty$ when $G\to\infty$ by \req{potential_int_scalar}. Therefore, combining with \req{potential_ineq}, the unique positive solution to the equation $U(1;G)=0$ exists on $(G_{conv}^*,\infty)$.
\eproof

Given \rthe{saturation_singleclass} and \rthe{upperboundbypotential}, one can improve the percolation threshold of a $(T, G, \Lambda(x), R, F)$-CPR system from $G_s^*$ to at least $G_{conv}^*$ by adopting the spatially-coupled method. Though the window size $w$ needs to be large to satisfy the condition in \req{unstable3333}, our numerical results in \rsec{irsaoneclass} show that adopting $w=2$ significantly improves the percolation threshold.

\rthe{upperboundbypotential} also reveals the existence of a numerical gap between $G_{conv}^*$ and $G_{up}^*$. However, as shown in \rsec{numerical}, this gap is small in most cases, with $|G_{up}^*-G_{conv}^*|$ typically being smaller than $10^{-2}$.

\bex{IRSA_regular}{(IRSA with a regular degree)}
In this example, we illustrate how $G_s^*$, $G_{conv}^*$ and $G_{up}^*$ are related for IRSA with a regular degree $d$, i.e., $\Lambda(x)=x^d$ and $P_\text{suc}(\rho)=e^{-\rho}$. Then we have from \req{udaloha} (with $D=1$) that
    \beq{IRSAp1111}
            U(p;G)=\frac{1}{d}\Big ((d-1)p^d-dp^{d-1}+\frac{1-e^{-Gdp^{d-1}}}{G}\Big ),
    \eeq
    and
    \beq{IRSAp2222}
        U'(p;G)=(d-1)p^{d-2}(p-1+e^{-Gdp^{d-1}}).
    \eeq
    For $d=3$, we use a computer search to find $G_s^*$, $G_{conv}^*$ and $G_{up}^*$ as follows:
    \bearn
        G_s^*&=&\sup\{G\in[0,1]|\\
        &&d(d-1)p^{d-2}(p-1+e^{-Gdp^{d-1}})>0,\forall p\in(0,1]\}\\
        &\approx&0.8184,\\
        G_{conv}^*&=&\sup\{G\in[0,1]|\\
        &&\min_{p\in[0,1]}(d-1)p^d-dp^{d-1}+\frac{1-e^{-Gdp^{d-1}}}{G}\geq0\}\\
        &\approx&0.9179,\\
        G_{up}^* &=&\{G: G=1-e^{-dG}\}\approx 0.9405.
    \eearn

\begin{figure}[!t]
        \centering
        \includegraphics[width=3.45in]{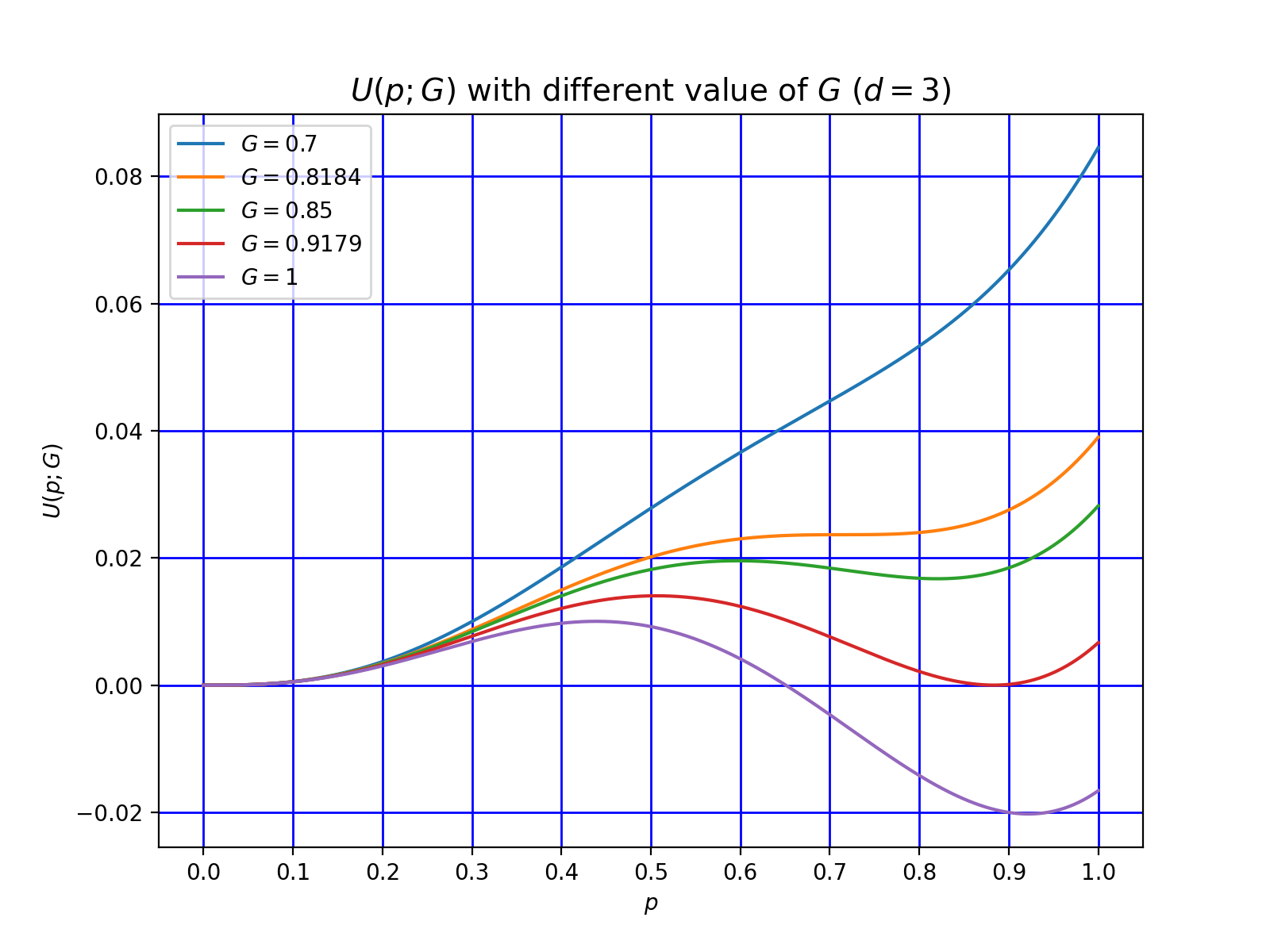}
        \caption{The plot of the potential function $U(p;G)$ as a function of $p$ for various values of $G$ in \rex{IRSA_regular} with $d=3$.}
        \label{fig:U_IRSA_regular}
    \end{figure}

     In \rfig{U_IRSA_regular}, we plot the potential function $U(p;G)$ as a function of $p$ for various values of $G$. This plot illustrates how the potential function changes for the parameter $p$ under various values of the parameter $G$.

\eex

\bsec{Numerical results}{numerical}

In this section, we provide numerical results for various systems of convolutional coded Poisson receivers. The convergence criteria are defined as $|q^{(i)}_{k,\ell}-q^{(i-1)}_{k,\ell}|<10^{-8}$ (or $|p^{(i)}_{k,\ell}-p^{(i-1)}_{k,\ell}|<10^{-8}$) for all $\ell$. The percolation threshold is determined by identifying the value of $G$ such that $\lim_{i\to\infty}p^{(i)}$ exhibits the first significant jump. The numerical results presented here are obtained using a step size of $\delta=0.0001$ for $G$ and $L=40$. All numerical values are rounded to four decimal places.

\bsubsec{Convolutional coded Poisson receivers with one class of users and one class of receivers}{irsaoneclass}

In this section, we focus on the single-class system (convolutional coded Poisson Receivers with $K=J=1$). This section aims to validate the results obtained in \rsec{saturation}.

In Table \ref{table:thrs_IRSA_regular}, we  provide percolation thresholds for the scalar admissible system ($w=1$) with $P_{{\rm suc}}(\rho)=e^{-\rho}$, i.e., IRSA.  These percolation thresholds are evaluated using the density evolution equation in \req{tag6666dmulm}. The columns $w=2,3,4$ present the percolation thresholds $G_{conv}^*(40,w)$ for $w=2,3,4$, respectively, evaluated using the equation in \req{tag6666dmulmc}. Additionally, we numerically evaluate $G_s^*$, $G_{conv}^*$, and $G_{up}^*$ by their definitions. As for the degree distribution, we set $\Lambda(x)=x^5$ to reproduce the numerical results presented in \cite{stefanovic2017asymptotic}. Table \ref{table:thrs_2foldALOHA} and Table \ref{table:thrs_3foldALOHA} show the corresponding results for the 2-fold ALOHA and the 3-fold ALOHA, respectively, with $P_{{\rm suc}}(\rho)$ being defined in \req{tfold1111}).

From these tables, we observe that $G_s^*<G_{conv}^*<G_{up}^*$ holds. Also, though the condition of the saturation theorem is not satisfied, the gap between $G_{conv}^*(L,w)$ and $G_{conv}^*$ is small. Specifically, the gap ranges from 0.0001 to 0.1. Note that $G_{conv}^*(L,w)$ and $G_{up}^*$ in the first two rows of each table match very well with those in Table I of \cite{Liva2012spatially} and Table I of \cite{stefanovic2017asymptotic}.

\begin{table}[!t]
\begin{center}
\caption{Percolation thresholds for convolutional IRSA.}
\label{table:thrs_IRSA_regular}
\resizebox{\columnwidth}{!}{\begin{tabular}{|cccccccc|}
\hline
\multicolumn{8}{|c|}{Convolutional IRSA} \\ \hline
\multicolumn{1}{|c|}{$d$}    & \multicolumn{1}{c|}{$w=1$}  & \multicolumn{1}{c|}{$w=2$}  & \multicolumn{1}{c|}{$w=3$}  & \multicolumn{1}{c|}{$w=4$}  & \multicolumn{1}{c|}{$G_s^*$} & \multicolumn{1}{c|}{$G_{conv}^*$}  & $G_{up}^*$ \\ \hline
\multicolumn{1}{|c|}{3} & \multicolumn{1}{c|}{0.8184} & \multicolumn{1}{c|}{0.9177} & \multicolumn{1}{c|}{0.9179} & \multicolumn{1}{c|}{0.9179} & \multicolumn{1}{c|}{0.8184}  & \multicolumn{1}{c|}{0.9179} & 0.9405    \\ \hline
\multicolumn{1}{|c|}{4} & \multicolumn{1}{c|}{0.7722} & \multicolumn{1}{c|}{0.9708} & \multicolumn{1}{c|}{0.9767} & \multicolumn{1}{c|}{0.9767} & \multicolumn{1}{c|}{0.7722}  & \multicolumn{1}{c|}{0.9767} & 0.9802    \\ \hline
\multicolumn{1}{|c|}{5} & \multicolumn{1}{c|}{0.7017} & \multicolumn{1}{c|}{0.9625} & \multicolumn{1}{c|}{0.9914} & \multicolumn{1}{c|}{0.9924} & \multicolumn{1}{c|}{0.7017}  & \multicolumn{1}{c|}{0.9924} & 0.9930    \\ \hline
\multicolumn{1}{|c|}{6} & \multicolumn{1}{c|}{0.6370} & \multicolumn{1}{c|}{0.9258} & \multicolumn{1}{c|}{0.9917} & \multicolumn{1}{c|}{0.9970} & \multicolumn{1}{c|}{0.6370}  & \multicolumn{1}{c|}{0.9973} & 0.9975    \\ \hline
\end{tabular}}
\end{center}
\end{table}

\begin{table}[!t]
\begin{center}
\caption{Percolation thresholds for the convolutional 2-fold ALOHA.}
\label{table:thrs_2foldALOHA}
\resizebox{\columnwidth}{!}{\begin{tabular}{|cccccccc|}
\hline
\multicolumn{8}{|c|}{Convolutional 2-fold ALOHA} \\ \hline
\multicolumn{1}{|c|}{$d$}    & \multicolumn{1}{c|}{$w=1$}  & \multicolumn{1}{c|}{$w=2$}  & \multicolumn{1}{c|}{$w=3$}  & \multicolumn{1}{c|}{$w=4$}  & \multicolumn{1}{c|}{$G_s^*$} & \multicolumn{1}{c|}{$G_{conv}^*$}  & $G_{up}^*$ \\ \hline
\multicolumn{1}{|c|}{3} & \multicolumn{1}{c|}{1.5528} & \multicolumn{1}{c|}{1.9560} & \multicolumn{1}{c|}{1.9760} & \multicolumn{1}{c|}{1.9763} & \multicolumn{1}{c|}{1.5528}  & \multicolumn{1}{c|}{1.9764} & 1.9790    \\ \hline
\multicolumn{1}{|c|}{4} & \multicolumn{1}{c|}{1.3336} & \multicolumn{1}{c|}{1.8966} & \multicolumn{1}{c|}{1.9894} & \multicolumn{1}{c|}{1.9961} & \multicolumn{1}{c|}{1.3336}  & \multicolumn{1}{c|}{1.9964} & 1.9966    \\ \hline
\multicolumn{1}{|c|}{5} & \multicolumn{1}{c|}{1.1577} & \multicolumn{1}{c|}{1.7722} & \multicolumn{1}{c|}{1.9639} & \multicolumn{1}{c|}{1.9955} & \multicolumn{1}{c|}{1.1577}  & \multicolumn{1}{c|}{1.9994} & 1.9995    \\ \hline
\multicolumn{1}{|c|}{6} & \multicolumn{1}{c|}{1.0216} & \multicolumn{1}{c|}{1.6326} & \multicolumn{1}{c|}{1.9071} & \multicolumn{1}{c|}{1.9833} & \multicolumn{1}{c|}{1.0216}  & \multicolumn{1}{c|}{1.9999} & 1.9999    \\ \hline
\end{tabular}}
\end{center}
\end{table}

\begin{table}[!t]
\begin{center}
\caption{Percolation thresholds for the convolutional $3$-fold ALOHA.}
\label{table:thrs_3foldALOHA}
\resizebox{\columnwidth}{!}{\begin{tabular}{|cccccccc|}
\hline
\multicolumn{8}{|c|}{Convolutional 3-fold ALOHA} \\ \hline
\multicolumn{1}{|c|}{$d$}    & \multicolumn{1}{c|}{$w=1$}  & \multicolumn{1}{c|}{$w=2$}  & \multicolumn{1}{c|}{$w=3$}  & \multicolumn{1}{c|}{$w=4$}  & \multicolumn{1}{c|}{$G_s^*$} & \multicolumn{1}{c|}{$G_{conv}^*$}  & $G_{up}^*$ \\ \hline
\multicolumn{1}{|c|}{3} & \multicolumn{1}{c|}{2.1744} & \multicolumn{1}{c|}{2.8990} & \multicolumn{1}{c|}{2.9874} & \multicolumn{1}{c|}{2.9916} & \multicolumn{1}{c|}{2.1744}  & \multicolumn{1}{c|}{2.9918} & 2.9923    \\ \hline
\multicolumn{1}{|c|}{4} & \multicolumn{1}{c|}{1.8108} & \multicolumn{1}{c|}{2.7127} & \multicolumn{1}{c|}{2.9577} & \multicolumn{1}{c|}{2.9950} & \multicolumn{1}{c|}{1.8108}  & \multicolumn{1}{c|}{2.9993} & 2.9994    \\ \hline
\multicolumn{1}{|c|}{5} & \multicolumn{1}{c|}{1.5456} & \multicolumn{1}{c|}{2.4744} & \multicolumn{1}{c|}{2.8636} & \multicolumn{1}{c|}{2.9739} & \multicolumn{1}{c|}{1.5456}  & \multicolumn{1}{c|}{2.9999} & 3.0000    \\ \hline
\multicolumn{1}{|c|}{6} & \multicolumn{1}{c|}{1.3487} & \multicolumn{1}{c|}{2.2425} & \multicolumn{1}{c|}{2.7240} & \multicolumn{1}{c|}{2.9213} & \multicolumn{1}{c|}{1.3487}  & \multicolumn{1}{c|}{2.9999} & 3.0000    \\ \hline
\end{tabular}}
\end{center}
\end{table}

\bsubsec{IRSA with two classes of users and two classes of receivers}{irsatwoclass}

In this section, we consider the IRSA system with two classes of users and two classes of receivers \rex{spatialTRSA_two_class} with the spatially coupled applied. As in \rex{spatialTRSA_two_class} and \cite{chang2022stability}, we consider the following two packet routing policies:

\begin{description}
    \item[1)] Complete sharing: every packet has an equal probability to be routed to the two classes of receivers, i.e., $r_{11}=r_{22}=r_{12}=r_{21}=0.5$.
    \item[2)] Receiver reservation: class 1 packets are routed to the two classes of receivers with an equal probability, i.e., $r_{11}=r_{12}=0.5$, and class 2 packets are routed to the class 2 receivers, i.e., $r_{21}=0$, $r_{22}=1$.
\end{description}

For our numerical validations, we set the number of iterations of \req{tag6666bmulr2} to $10,000$. We set the number of stages $L=40$ and the window size $w$ is set from 2 to 4. These numerical results are obtained from a grid search with a step size of $\delta=0.01$ for both $G_1$ and $G_2$. For the sake of numerical stability in our computation, we round up $p^{(10000)}_{k,L/2}(G_1,G_2)$ to 0 if its computed value is smaller than $10^{-5}$.

\begin{figure}[!t]
    \centering
    \includegraphics[width=3.45in]{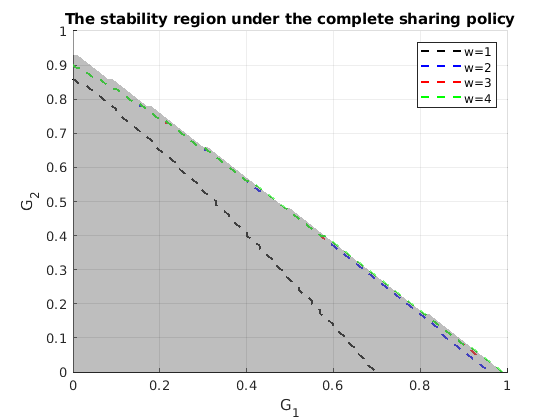}
    \caption{The boundaries of the stability region of the convolutional IRSA system with two classes of users and two classes of receivers under the complete sharing policy. The gray area presents the outer bound \req{IRSA2class2222} for the complete sharing policy.}
    \label{fig:class1}
\end{figure}

\begin{figure}[!t]
    \centering
    \includegraphics[width=3.45in]{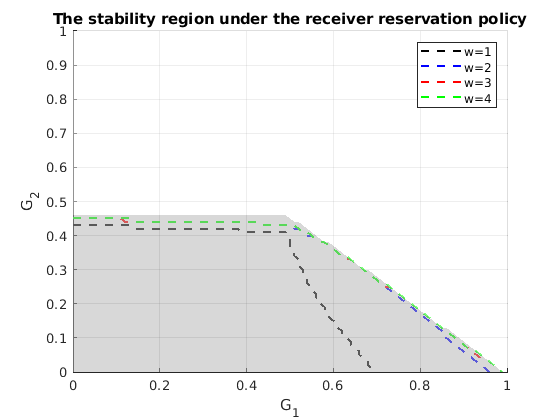}
    \caption{The boundaries of the stability region of the convolutional IRSA system with two classes of users and two classes of receivers under the receiver reservation policy. The gray area is the intersection of the outer bounds \req{IRSA2clas5555} and \req{IRSA2class9999} for the receiver reservation policy.}
    \label{fig:class2}
\end{figure}

In \rfig{class1} and \rfig{class2}, we depict the stability region boundaries for two classes of users with two different degree distributions for $w=2,3,4$. We choose $\Lambda_1(x)=x^5$ and $\Lambda_2(x)=0.5102x^2+0.4898x^4$, where $\Lambda_2(x)$ is selected from Table 1 of \cite{liva2011graph} to achieve a high percolation threshold of 0.868 in IRSA with a single class of users. The legend $w=1$ denotes the conventional coded Poisson receiver (without convolution), reproducing the numerical results in \cite{chang2022stability}. The colored dashed lines represent the boundary of stability regions for different values of $w$. In both figures, the boundaries of the stability regions almost overlap for $w=2,3,4$.

Furthermore, the shaded regions in both figures represent the outer bounds of the stability region evaluated in \rex{spatialTRSA_two_class}. Moreover, in both figures, we observe a significant enlargement of the stability region from $w=1$ to $w=2$, consistent with the findings of \rthe{saturation}. Therefore, the spatial coupling effect can enlarge the stability region. This expansion continues as $w$ monotonically increases from 2 to 4. Eventually, the boundaries of the expanded stability regions are close to their outer bounds. These empirical results align with the single-class scenario discussed in \rsubsec{irsaoneclass}. For \rfig{class2}, the reservation policy ($r_{21}=0$) notably constrains the outer bound \req{IRSA2class9999} for $G_2$. Thus, the stability region is significantly limited in $G_2$ compared to that of $G_1$.

\bsec{Conclusion}{con}

In this paper, we introduced a probabilistic framework known as convolutional coded Poisson receivers, extending the concept of coded Poisson receivers by drawing inspiration from convolutional LDPC codes. The main contributions of this paper are summarized as follows:

\begin{description}
\item[(i)] We established the outer bounds for the stability region of a system of CPRs in \rthe{capacity}, applicable to both CPRs and CCPRs. These outer bounds are extended to encompass multiple traffic classes.
\item[(ii)] We established the saturation theorem (\rthe{saturation_singleclass}) as a sufficient condition for CCPRs to demonstrate a higher percolation threshold compared to conventional CPRs. We validated the numerical results in \rsec{irsaoneclass}.
\item[(iii)] For the single-class user and receiver scenario of CCPRs, we employed the potential function to characterize three important thresholds, namely $G_s^*<G_{conv}^*<G_{up}^*$, as presented in \rthe{upperboundbypotential}.
\item[(iv)] Our numerical results showed that the stability region of CCPRs might approach its outer bounds under finite iterations of the density evolution equations.
\end{description}

As a future direction, we are interested in exploring the parallel statement to Theorem 10 in \cite{kudekar2011threshold}. Specifically, we aim to investigate whether
$$\lim_{w\to\infty}\lim_{L\to\infty}G_{conv}^*(L,w)=G_{conv}^*.$$



\begin{IEEEbiography}[{\includegraphics[width=1in,height=1.25in,clip,keepaspectratio]{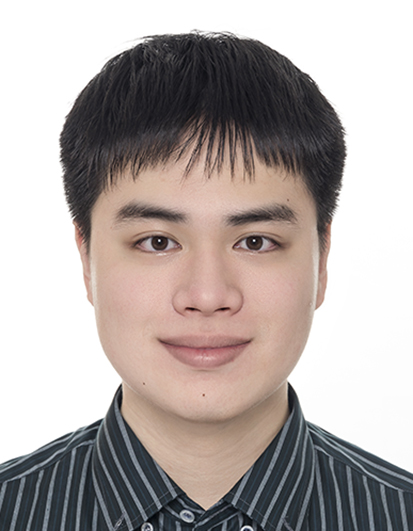}}]
{Cheng-En Lee} received the B.S. degree in electrical engineering from National Tsing Hua University, Hsinchu, Taiwan in 2021. He is a graduate student at the Institute of Information Systems and Applications, National Tsing Hua University, Hsinchu, Taiwan, in 2023. His research interests are in network science, deep learning, and wireless communications.
\end{IEEEbiography}

\begin{IEEEbiography}[{\includegraphics[width=1in,height=1.25in,clip,keepaspectratio]{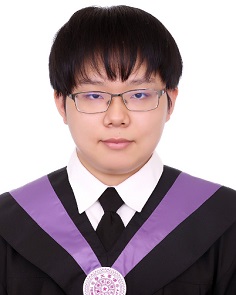}}]
	{Kuo-Yu Liao}
received the B.S. degree in mathematics in 2022 from National Tsing Hua University. He is currently pursuing the M.S. degree in the Institute of Communications Engineering, National Tsing Hua University, Hsinchu, Taiwan.
His research interest is in 5G and beyond wireless communication.
\end{IEEEbiography}

\begin{IEEEbiography}[{\includegraphics[width=1in,height=1.25in,clip,keepaspectratio]{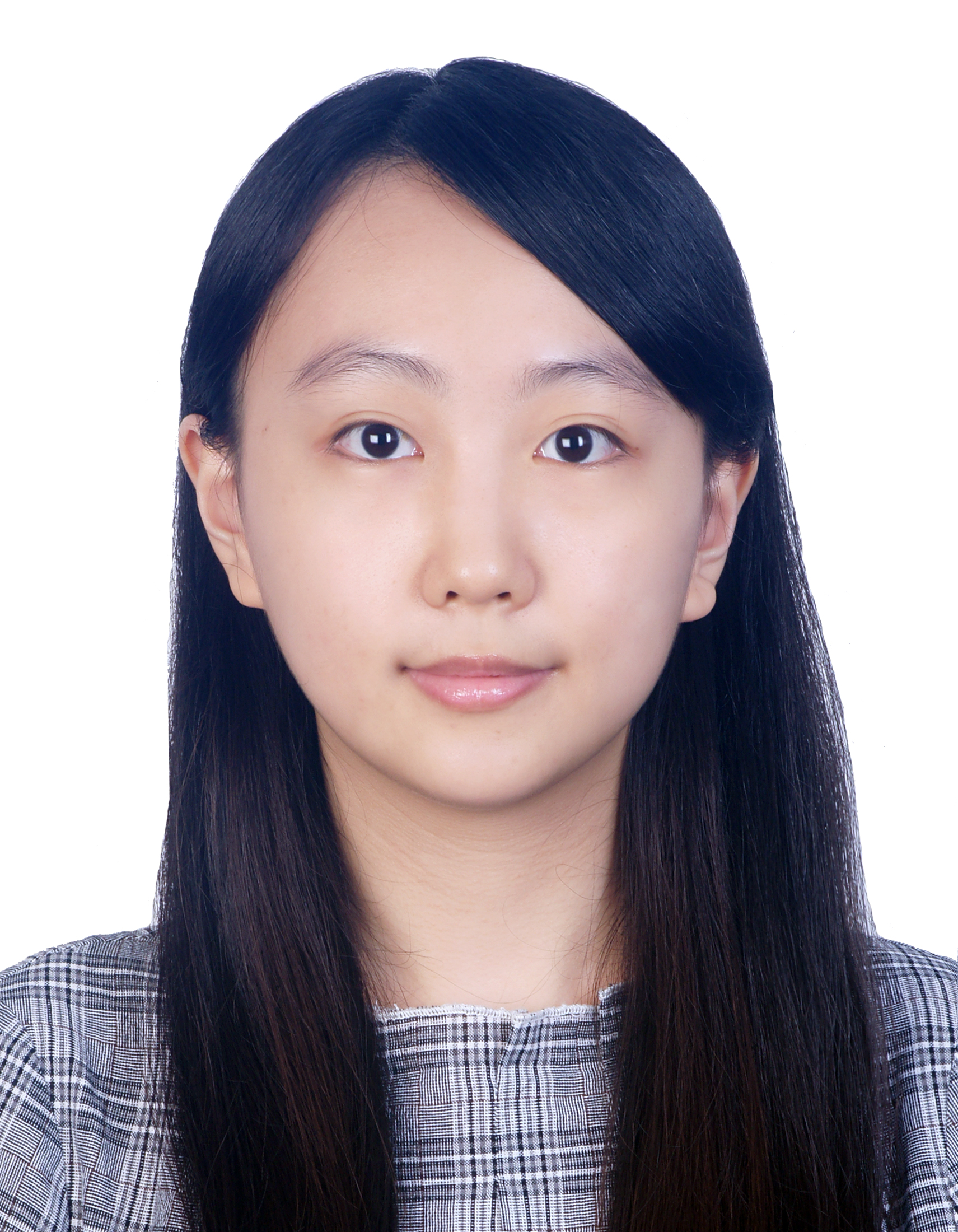}}]
{Hsiao-Wen Yu} received the B.S. degree in electrical engineering from Chang Gung University, Taoyuan, Taiwan, in 2021. She is currently pursuing the M.S. degree in the Institute of Communications Engineering, National Tsing Hua University, Hsinchu, Taiwan. Her research interest is in 5G and beyond wireless communication.
\end{IEEEbiography}

\begin{IEEEbiography}[{\includegraphics[width=1in,height=1.25in,clip,keepaspectratio]{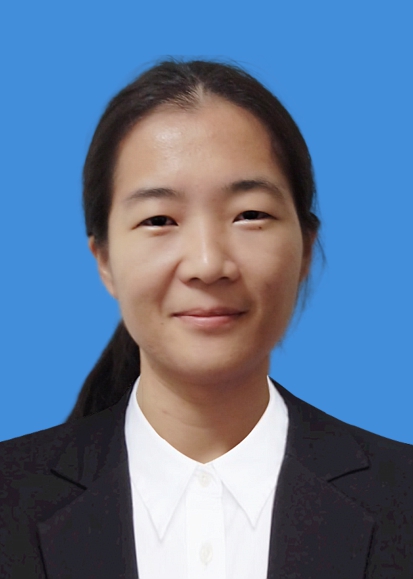}}]
{Ruhui Zhang} received the B.S. degree in advertisement from Xiamen University, Xiamen, China, in 2014, and the Ph.D. degree with the Department of Computer Science, National Tsing Hua University, Hsinchu, Taiwan, in 2023. Her research interests are in network science, data analytics, and wireless communications.
\end{IEEEbiography}

\begin{IEEEbiography}[{\includegraphics[width=1in,height=1.25in,clip,keepaspectratio]{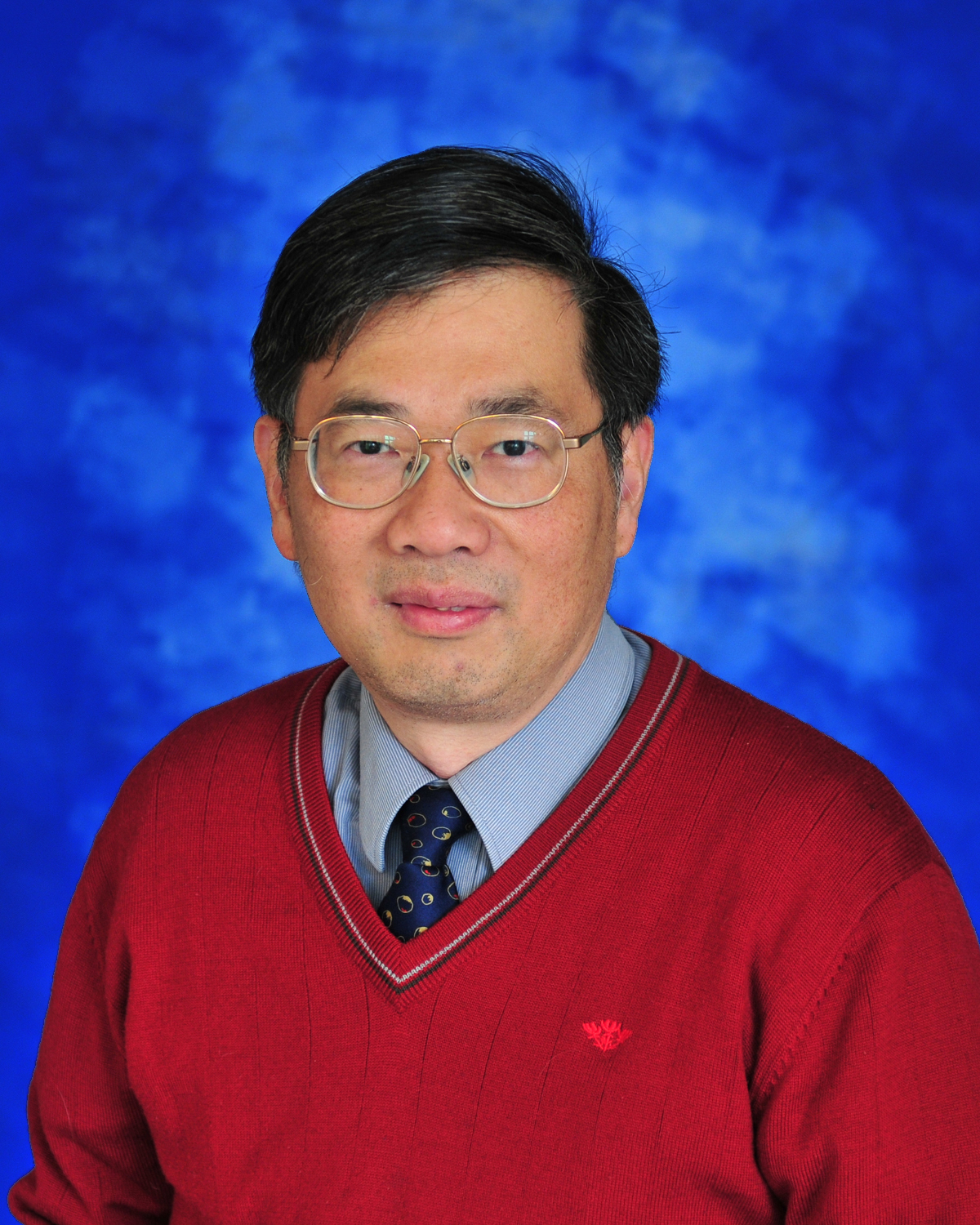}}]
{Cheng-Shang Chang}(S'85-M'86-M'89-SM'93-F'04) received the B.S. degree from National Taiwan University, Taipei, Taiwan, in 1983, and the M.S. and Ph.D. degrees from Columbia University, New York, NY, USA, in 1986 and 1989, respectively, all in electrical engineering.
	
From 1989 to 1993, he was employed as a Research Staff Member with the IBM Thomas J. Watson Research Center, Yorktown Heights, NY, USA. Since 1993, he has been with the Department of Electrical Engineering, National Tsing Hua University, Taiwan, where he is a Tsing Hua Distinguished Chair Professor. He is the author of the book Performance Guarantees in Communication Networks (Springer, 2000) and the coauthor of the book Principles, Architectures and Mathematical Theory of High-Performance Packet Switches (Ministry of Education, R.O.C., 2006). His current research interests are concerned with network science, big data analytics, mathematical modeling of the Internet, and high-speed switching.
	
Dr. Chang served as an Editor for Operations Research from 1992 to 1999, an Editor for the {\em IEEE/ACM TRANSACTIONS ON NETWORKING} from 2007 to 2009, and an Editor for the {\em IEEE TRANSACTIONS ON NETWORK SCIENCE AND ENGINEERING} from 2014 to 2017. He is currently serving as an Editor-at-Large for the {\em IEEE/ACM TRANSACTIONS ON NETWORKING}. He is a member of IFIP Working Group 7.3. He received an IBM Outstanding Innovation Award in 1992, an IBM Faculty Partnership Award in 2001, and Outstanding Research Awards from the National Science Council, Taiwan, in 1998, 2000, and 2002, respectively. He also received Outstanding Teaching Awards from both the College of EECS and the university itself in 2003. He was appointed as the first Y. Z. Hsu Scientific Chair Professor in 2002. He received the Merit NSC Research Fellow Award from the National Science Council, R.O.C. in 2011. He also received the Academic Award in 2011 and the National Chair Professorship in 2017 from the Ministry of Education, R.O.C. He is the recipient of the 2017 IEEE INFOCOM Achievement Award.
\end{IEEEbiography}

\begin{IEEEbiography}[{\includegraphics[width=1in,height=1.25in,clip,keepaspectratio]{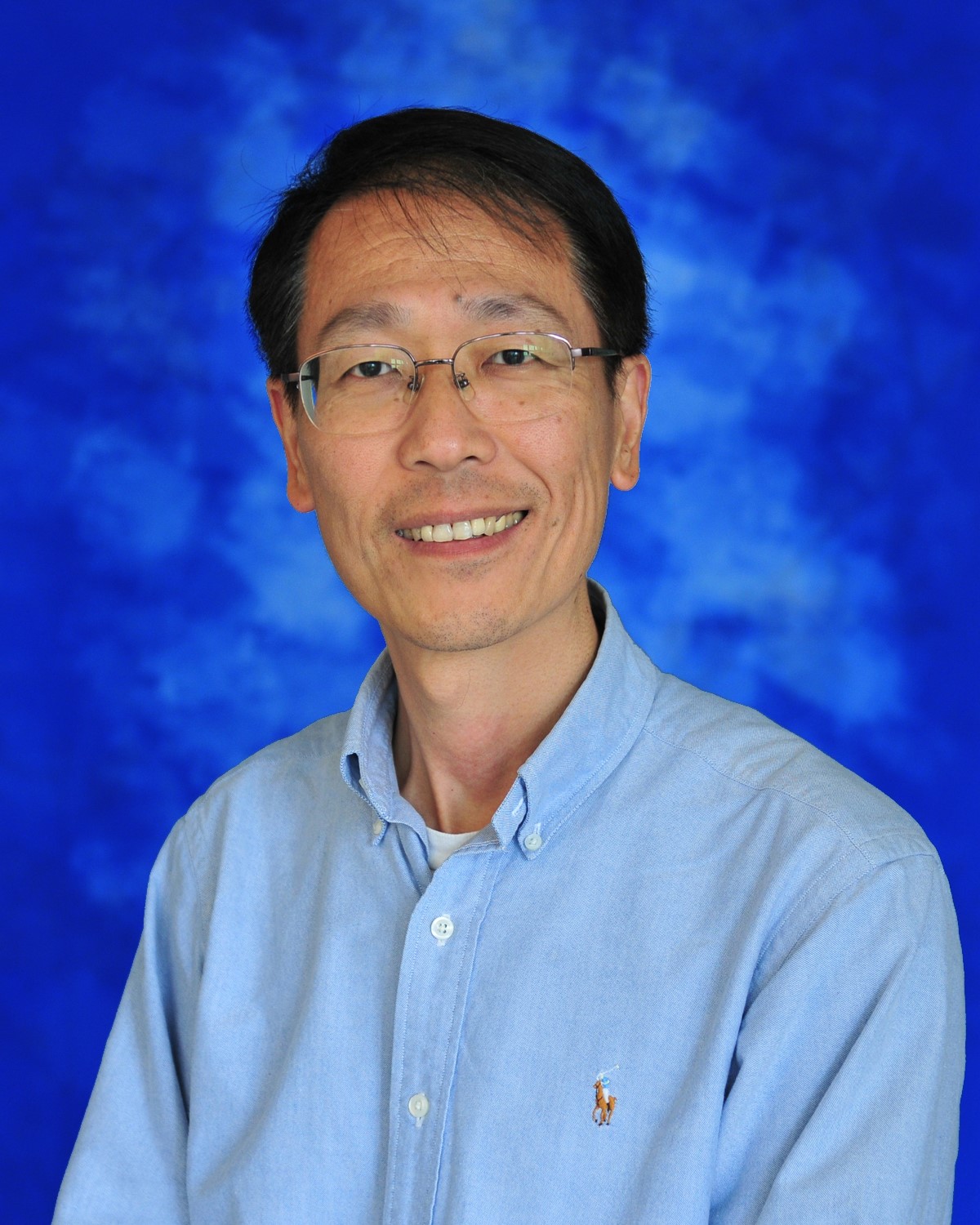}}]
{Duan-Shin Lee}(S'89-M'90-SM'98) received the B.S. degree from National Tsing Hua University, Taiwan, in 1983, and the MS and Ph.D. degrees from Columbia University, New York, in 1987 and 1990, all in electrical engineering. He worked as a research staff member at the C\&C Research Laboratory of NEC USA, Inc. in Princeton, New Jersey from 1990 to 1998. He joined the Department of Computer Science of National Tsing Hua University in Hsinchu, Taiwan, in 1998. Since August 2003, he has been a professor. He received a best paper award from the Y.Z. Hsu Foundation in 2006. He served as an editor for the Journal of Information Science and Engineering between 2013 and 2015.  He is currently an editor for Performance Evaluation. Dr. Lee's current research interests are network science, game theory, machine learning and high-speed networks. He is a senior IEEE member.
\end{IEEEbiography}

\clearpage

\appendices

\section{Proof of \rlem{maxdecode}}\label{proof_of_maxdecode}

\bproof
Consider a particular Poisson receiver. During the SIC decoding process, we assume that there are $n_{k_\ell}>0$ class $k_\ell$ packets that are decoded during the $i_\ell$-th iteration, $\ell=1,2, \ldots, M$ for some $M$. Without loss of generality, we assume that $i_1 \le i_2 \le \ldots \le i_M$. Let ${\bf e}_k$ be the $1 \times K$ vector with its $k$-th element being 1 and 0 otherwise. Thus, the number of packets decoded by this receiver can be represented by the vector $\sum_{\ell=1}^M n_{k_\ell} {\bf e}_{k_\ell}$. To prove this lemma, we need to show that the decoded packets by this Poisson receiver is within the capacity region of the $\phi$-ALOHA receiver, i.e.,
\beq{maxdecode1111}
    \phi(\sum_{\ell=1}^M n_{k_\ell} {\bf e}_{k_\ell})=\sum_{\ell=1}^M n_{k_\ell} {\bf e}_{k_\ell}
\eeq
This is equivalent to showing that
\beq{maxdecode2222}
    \phi^c_{k_\ell}(\sum_{\ell=1}^M n_{k_\ell} {\bf e}_{k_\ell})=0,
\eeq
for $\ell=1,2, \ldots, M$. From the all-or-nothing property, we know that $M \le K$ as packets of the same class are decoded in the same iteration.

We prove \req{maxdecode2222} by induction. Suppose that there are $n^{(i)}=(n_1^{(i)}, n_2^{(i)}, \ldots, n_K^{(i)})$ remaining packets at the receiver during the $i$-th iteration. As the number of decoded class $k$ packets cannot be larger than the number of remaining class $k$ packets, we have
\beq{maxdecode2255}
    n_{k_1} {\bf e}_{k_1} \le n^{(i_1)}.
\eeq

As there are $n_{k_1}>0$ class $k_1$ packets that are decoded during the $i_1$-th iteration, we have from the monotone property and the all-or-nothing property that
\beq{maxdeonde3333}
    \phi^c_{k_1}(\sum_{\ell=1}^M n_{k_\ell} {\bf e}_{k_\ell})\le \phi^c_{k_1}(n^{(i_1)})=0.
\eeq
Now assume that \req{maxdecode2222} holds for $\ell=1,2,\ldots, m$ as the induction hypothesis. From the induction hypothesis, we have
\beq{maxdecode4444}
    \phi^c(\sum_{\ell=1}^M n_{k_\ell} {\bf e}_{k_\ell})\le \sum_{\ell=m+1}^M n_{k_\ell} {\bf e}_{k_\ell}.
\eeq
From the closure property, the monotone property and \req{maxdecode4444}, we know that
\bear{maxdecode5555}
    &&\phi^c_{k_{m+1}}(\sum_{\ell=1}^M n_{k_\ell} {\bf e}_{k_\ell}) =\phi^c_{k_{m+1}}(\phi^c(\sum_{\ell=1}^M n_{k_\ell} {\bf e}_{k_\ell})) \nonumber\\
    &&\le \phi^c_{k_{m+1}}(\sum_{\ell=m+1}^M n_{k_\ell} {\bf e}_{k_\ell}).
\eear
As the number of decoded class $k$ packets cannot be larger than the number of remaining class $k$ packets, we have at the $i_{m+1}$-th iteration that
\beq{maxdecode2255b}
    \sum_{\ell=m+1}^M n_{k_\ell} {\bf e}_{k_\ell}\le n^{(i_{m+1})}.
\eeq
It follows from the monotone property and the all-or-nothing property that
\beq{maxdecode6666}
    \phi^c_{k_{m+1}}(\sum_{\ell=m+1}^M n_{k_\ell} {\bf e}_{k_\ell})\le \phi^c_{k_{m+1}}(n^{(i_{m+1})})=0.
\eeq
Thus, $\phi^c_{k_{m+1}}(\sum_{\ell=1}^M n_{k_\ell} {\bf e}_{k_\ell})=0$, and we complete the induction.
\eproof

\section{Density evolution for the circular convolutional CPR in \rcor{mainextc}}\label{proof_of_mainextc}

In this appendix, we conduct the density evolution analysis for the circular convolutional CPR with $L$ stages. The analysis is similar to that in \cite{chang2022stability} that uses the density evolution method in \cite{luby1998analysis,luby1998analysisb,richardson2001capacity,liva2011graph,paolini2011graph,paolini2012random} and the reduced Poisson offered load argument in \cite{kelly2011reversibility,walrand1983probabilistic,kelly1991loss,chang2020Poisson}. For our analysis, we call an edge a class $k$ edge if the user end of the edge is connected to a class $k$ user. Also, we call an edge a class $(k,j)$-edge if the receiver (resp. user) end of the edge is connected to a class $j$ receiver (resp. class $k$ user).

The density evolution analysis consists of the following steps:

\noindent (i) The initial offered load of class $k$ packets to a class $j$ Poisson receiver in the $\ell^{th}$ stage, defined as the expected number of class $k$ packets transmitted to that receiver, is
\beq{mean4444mulr}
    \rho_{k,j,\ell}= \sum_{\hat \ell=\ell \ominus (w-1)}^\ell G_{k}^{(\hat \ell)} \Lambda_{k}^\prime(1) \frac{1}{w} r_{k,j}/F_j.
\eeq

To see \req{mean4444mulr}, note that (a) there are $G_k^{(\hat \ell)} T$ class $k$ users in the $\hat \ell^{th}$ stage, (b) each class $k$ user transmits on average $\Lambda_{k}^\prime(1)$ copies, (c) each copy in the $\hat \ell^{th}$ stage (with $\hat \ell \in [\ell \ominus (w-1),\ell]$), is sent to  class $j$ Poisson receivers in the $\ell^{th}$ stage with the routing probability $r_{k,j}/w$, and (d) a copy sent to class $j$ Poisson receivers is uniformly distributed among the $F_j T$ class $j$ Poisson receivers. When $T$ goes to infinity, the number of class $k$ packets at a class $j$ receiver in the $\ell^{th}$ stage converges (from a binomial random variable) to a Poisson random variable with mean $\rho_{k,j,\ell}$, and the degree distribution of class $k$ packets at a class $j$ receiver node is a Poisson distribution with mean $\rho_{k,j,\ell}$.

\noindent (ii) Let $q_{k,\ell}^{(i)}$ be the probability that the {\em user end} of a randomly selected class $k$ edge in the $\ell^{th}$ stage has not been successfully received after the $i^{th}$ SIC iteration. The offered load of class $k$ packets to a class $j$ Poisson receiver in the $\ell^{th}$ stage after the $i^{th}$ SIC iteration has a Poisson distribution with mean $\rho_{k,j,\ell}^{(i)}$ (from the reduced offered load argument), where
\beq{mean4444mulrb}
    \rho_{k,j,\ell}^{(i)}= \sum_{\hat \ell=\ell \ominus (w-1)}^\ell q_{k,\hat \ell}^{(i)} G_{k}^{(\hat \ell)} \Lambda_{k}^\prime(1) \frac{1}{w} r_{k,j}/F_j.
\eeq
Let
\beq{rho0000r1}
    \trhojl^{(i)}=(\rho_{1,j,\ell}^{(i)}, \rho_{2,j,\ell}^{(i)}, \ldots, \rho_{K,j,\ell}^{(i)}).
\eeq
Note that one can represent the offered load at a class $j$ Poisson receiver in the $\ell^{th}$ stage after the $i^{th}$ SIC iteration by the vector $\trhojl^{(i)}$.

\noindent (iii) Let $p_{k,j,\ell}^{(i+1)}$ be the probability that the {\em receiver end} of a randomly selected class $(k,j)$-edge in the $\ell^{th}$ stage has not been successfully received after the $(i+1)^{th}$ SIC iteration. Then
\beq{tag6666amulr}
    p_{k,j,\ell}^{(i+1)}=1- P_{{\rm suc},k,j}(\trhojl^{(i)}).
\eeq
That \req{tag6666amulr} holds follows directly from the definition of a Poisson receiver in \rdef{Poissonmul} as the offered load at a class $j$ Poisson receiver in the $\ell^{th}$ stage after the $i^{th}$ SIC iteration is $\trhojl^{(i)}$.

\noindent (iv) Let $p_{k,\ell}^{(i+1)}$ be the probability that  the {\em receiver end} of a randomly selected class $k$ edge in the $\ell^{th}$ stage has not been successfully received after the $(i+1)^{th}$ SIC iteration. Since a class $k$ edge in the $\ell^{th}$ stage is a class $(k,j)$-edge in the $\hat \ell^{th}$ stage (for $\hat \ell \in [\ell,\ell\oplus (w-1)]$)  with probability $r_{k,j}/w$, it follows that
\bear{tag6666bmulr}
    p_{k,\ell}^{(i+1)}&=&\sum_{\hat \ell=\ell}^{\ell\oplus (w-1)}\sum_{j=1}^J \frac{1}{w} r_{k,j}p_{k,j,\hat \ell}^{(i+1)} \nonumber\\
    &=&1-\sum_{\hat \ell=\ell}^{\ell\oplus (w-1)}\sum_{j=1}^J \frac{1}{w} r_{k,j}P_{{\rm suc},k,j}(\trhojhl^{(i)}).\nonumber\\
\eear

\noindent (v) The probability $q_{k,\ell}^{(i)}$ can be computed recursively from the following equation:
\beq{tag6666cmulr}
    q_{k,\ell}^{(i+1)}=\lambda_{k}(1- \sum_{\hat \ell=\ell}^{\ell\oplus (w-1)}\sum_{j=1}^J \frac{1}{w} r_{k,j}P_{{\rm suc},k,j}(\trhojhl^{(i)})),
\eeq
with $q_{k,\ell}^{(0)}=1$. To see this, note that a packet sent from a user (the user end of the bipartite graph) can be successfully received if at least one of its copies is successfully received at the {\em receiver} end. Since the probability that the user end of a randomly selected class $k$ edge  has additional $d$ edges is $\lambda_{k,d}$, the probability that the {\em user} end of a randomly selected class $k$ edge  cannot be successfully received after the $(i+1)^{th}$ iteration is
\bear{tag2222mulr}
    q_{k,\ell}^{(i+1)}&=&1-\sum_{d=0}^\infty \lambda_{k,d} \cdot \Big (1-(p_{k,\ell}^{(i+1)})^{d} \Big) \nonumber\\
    &=&\lambda_{k}(p_{k,\ell}^{(i+1)}).
\eear
Using \req{tag6666bmulr} in \req{tag2222mulr} yields \req{tag6666cmulr}.

\noindent (vi) Let $\tilde P_{{\rm suc},k,\ell}^{(i)}$ be the probability that a packet sent from a randomly selected {\em class $k$ user} in the $\ell^{th}$ stage can be successfully received after the $i^{th}$ iteration. Such a probability is the probability that at least one copy of the packet has been successfully received after the $i^{th}$ iteration. Since the probability that a randomly selected {\em class $k$ user} has  $d$ edges is $\Lambda_{k,d}$, we have from \req{tag6666bmulr} that
\bear{mean5555dmulr}
    &&\tilde P_{{\rm suc},k,\ell}^{(i)}\nonumber\\
    &&=\sum_{d=0}^\infty \Lambda_{k,d} \cdot \Big (1-(p_{k,\ell}^{(i)})^{d} \Big)\nonumber\\
    &&=1-\Lambda_k\Big (1- \sum_{\hat \ell=\ell}^{\ell\oplus (w-1)}\sum_{j=1}^J \frac{1}{w} r_{k,j}P_{{\rm suc},k,j}(\trhojhl^{(i-1)})\Big). \nonumber \\
\eear
\eproof

\section{Proof of \rthe{saturation_singleclass}}\label{proof_of_saturation_singleclass}

The proof of \rthe{saturation_singleclass} requires several lemmas similar to those in \cite{yedla2012simple} using the notations employed in this paper. Though these lemmas may not have a direct physical interpretation, they are crucial for establishing the proof of \rthe{saturation_singleclass}.

Let $\tilde{L}=L-w+1$. We first reverse the indices of the recursion equation \req{simplified}. Let $\tilde{p}_\ell^{(i)}=p_{{\tilde{L}}-\ell+1}^{(i)}$, \req{simplified} becomes
\bear{reversed1111}
    &&\tilde{p}_\ell^{(i+1)}=1-\sum_{\hat{\ell}=\tilde{L}-\ell+1}^{\tilde{L}-\ell+w}\frac{1}{w}P_{\text{suc}}\bigg(\frac{G\Lambda^\prime(1)}{w}\nonumber\\
    &&\sum_{\tilde \ell=\max[1,\hat{\ell}-w+1]}^{\min[\tilde{L},\hat{\ell}]}\lambda(\tilde{p}_{\tilde{L}-\tilde \ell+1}^{(i)})\bigg).\nonumber\\
\eear

Let $\tilde \ell^\prime={\tilde{L}}-\tilde \ell+1$. The upper and lower indices of the second summation could be changed into
$$\tilde \ell^\prime=\min[\tilde{L},\tilde{L}-\hat\ell+w]$$
and
$$\tilde \ell^\prime=\max[1,\tilde{L}-\hat\ell+1].$$
Note that the latter becomes not larger than the former in this case. Thus, \req{reversed1111} further becomes
\beq{reversed3333}
    \tilde{p}_\ell^{(i+1)}=1-\sum_{\hat{\ell}=\tilde{L}+1-\ell}^{\tilde{L}+w-\ell}\frac{1}{w}P_{\text{suc}}\bigg(\frac{G\Lambda^\prime(1)}{w}\sum_{\tilde \ell^\prime=\max[1,\tilde{L} - \hat{\ell} + 1]}^{\min[\tilde{L},\tilde{L} - \hat{\ell} + w]}\lambda(\tilde{p}_{\tilde \ell^\prime}^{(i)})\bigg).
\eeq

Hence, reversing the indices, the convolutional $(T, G, \Lambda(x), R, F,w)$-CPR with $L$ stages with one class of users is governed by the following recursion equation:
\beq{reversed2222}
    p_\ell^{(i+1)}=1-\sum_{\hat{\ell}=\tilde{L}+1-\ell}^{\tilde{L}+w-\ell}\frac{1}{w}P_{\text{suc}}\bigg(\frac{G\Lambda^\prime(1)}{w}\sum_{\tilde \ell=\max[1,\tilde{L} - \hat{\ell} + 1]}^{\min[\tilde{L},\tilde{L} - \hat{\ell} + w]}\lambda(p_{\tilde \ell}^{(i)})\bigg),
\eeq
where the initial condition is $\mathbf{p}^{(0)}=(1, 1, \ldots, 1)$.

A $(T,G, \Lambda(x), R, F)$-CPR is a kind of scalar admissible system characterized by a pair of functions $(f,h)$, which can be represented as a bipartite graph. If we construct a large bipartite graph as the construction of convolutional $(T, G, \Lambda(x), R, F,w)$-CPR systems in \rsec{CCPR}, we obtain a {\em basic spatially-coupled} system in \cite{yedla2012simple}, which is also parametrized by a pair of functions $(f,h)$.

\bdefin{basicSCsystem}{(\bf The basic spatially-coupled system} (cf. Def. 10 in \cite{yedla2012simple})
    The basic spatially-coupled system is defined by concatenating $L$ bipartite graphs of a scalar admissible system. The edges in the concatenated bipartite graphs are rewired to form a single bipartite graph as follows: The receiver end of each edge in the $\ell$-th copy is rewired to the corresponding receiver node in the $\hat{\ell}$-th copy, where $\hat{\ell}$ is chosen uniformly in $[\ell, \ell\oplus (w-1)]$. Then, we reverse the index. Additionally, the recursion is defined as:
    \bear{basicSC}
        p_{\ell}^{(i+1)}=\frac{1}{w}\sum_{\hat{\ell}=\tilde{L}+1-\ell}^{\tilde{L}+w-\ell}f\bigg(\frac{1}{w}\sum_{\tilde \ell=\max[1,\tilde{L} - \hat{\ell} + 1]}^{\min[\tilde{L},\tilde{L} - \hat{\ell} + w]}h(p_{\tilde \ell}^{(i)});G\bigg),\nonumber\\
    \eear
    for $\ell=1,2,\ldots,L-w+1$. Also, conditions (P1)-(P4) in \rdef{scalar_admissible_system} should be satisfied.

    Define $\mathbf{h}(\mathbf{p})=(h(p_1), h(p_2),\ldots,h(p_{\tilde{L}}))$ and $\mathbf{f}(\mathbf{p};G)=(f(p_1;G), f(p_2;G),\ldots,f(p_{\tilde{L}};G))$, then the {\em vector recursion} of \req{basicSC} is given by:
    \beq{basicSCvector}
        \mathbf{p}^{(i+1)}=\mathbf{A}_2\mathbf{f}(\mathbf{A}_2^T\mathbf{h}(\mathbf{p}^{(i)});G),
    \eeq
    where $\mathbf{A}_2$ is an $\tilde{L}\times L$ matrix defined as
    \beq{matrixA2}
        \mathbf{A}_2=\frac{1}{w}\begin{bmatrix}
        1      & 1      & \ldots  & 1      & 0      & 0      & \ldots  & 0      \\
        0      & 1      & 1      & \ldots  & 1      & 0      & \ldots  & 0      \\
        \vdots & \ddots & \ddots & \ddots & \ddots & \ddots & \ddots & \vdots \\
        0      & \ldots  & 0      & 1      & 1      & \ldots  & 1      & 0      \\
        0      & \ldots  & 0      & 0      & 1      & 1      & \ldots  & 1      \\
        \end{bmatrix},
    \eeq
    and all the columns of $\mathbf{A}$ contain $w$ ones.
\edefin

Next, by examining the conditions in \rdef{basicSCsystem}, we prove that CCPRs are one kind of basic spatially-coupled system.
\blem{estimate}
    The convolutional $(T, G, \Lambda(x), R, F,w)$-CPR system is a basic spatially-coupled system. That is, for $\ell=1,2,\ldots, L-w+1$,
    \bear{CCPRupperbound}
        p_{\ell}^{(i+1)}=\frac{1}{w}\sum_{\hat{\ell}=\tilde{L}+1-\ell}^{\tilde{L}+w-\ell} \bigg(1 -P_{\text{suc}}\Big(\frac{G\Lambda^\prime(1)}{w}\sum_{\tilde \ell=\max[1,\tilde{L} - \hat{\ell} + 1]}^{\min[\tilde{L},\tilde{L} - \hat{\ell} + w]}\lambda(p_{\tilde \ell}^{(i)})\Big)\bigg).\nonumber\\
    \eear
\elem

\bproof
First, rearranging \req{reversed2222},
\bear{pfapp1}
    &&p_{\ell}^{(i+1)}=1-\sum_{\hat{\ell}=\tilde{L}+1-\ell}^{\tilde{L}+w-\ell}\frac{1}{w}P_{\text{suc}}\nonumber\\
    &&\bigg(\frac{G\Lambda^\prime(1)}{w}\sum_{\tilde \ell=\max[1,\tilde{L} - \hat{\ell} + 1]}^{\min[\tilde{L},\tilde{L} - \hat{\ell} + w]}\lambda(p_{\tilde \ell}^{(i)})\bigg).\nonumber\\
\eear

Then, insert 1 into the summation,
\bear{pfapp2}
    &&p_\ell^{(i+1)}=\sum_{\hat{\ell}=\tilde{L}+1-\ell}^{\tilde{L}+w-\ell}\bigg(\frac{1}{w}-\nonumber\\
    &&\frac{1}{w}P_\text{suc}\bigg(\frac{G\Lambda^\prime(1)}{w}\sum_{\tilde \ell=\max[1,\tilde{L} - \hat{\ell} + 1]}^{\min[\tilde{L},\tilde{L} - \hat{\ell} + w]}\lambda(p_{\tilde \ell}^{(i)})\bigg)\bigg).\nonumber\\
\eear

Taking $1/w$ out from the summation,
\bear{pfapp3}
    &&p_\ell^{(i+1)}=\frac{1}{w}\sum_{\hat{\ell}=\tilde{L}+1-\ell}^{\tilde{L}+w-\ell}\bigg(1-\nonumber\\
    &&P_\text{suc}\bigg(\frac{G\Lambda^\prime(1)}{w}\sum_{\tilde \ell=\max[1,\tilde{L} - \hat{\ell} + 1]}^{\min[\tilde{L},\tilde{L} - \hat{\ell} + w]}\lambda(p_{\tilde \ell}^{(i)})\bigg)\bigg).\nonumber\\
\eear

Assign
\begin{align}
    f(p;G)&=1-P_\text{suc}(pG\Lambda^\prime(1)),\\
    h(p)&=\lambda(p).
\end{align}

By \rlem{CPR_SA}, conditions (1)-(4) of \rdef{scalar_admissible_system} are satisfied.
\eproof

Now we introduce the {\em vector one-sided spatially-coupled recursion} system (briefly, a one-sided system in this paper) in \cite{yedla2012simple} and \cite{Yedla2012vector}. It is also parametrized by a pair of functions $(f,h)$. Hence, there exists a correspondence between one-sided systems, basic spatially-coupled systems, and scalar admissible systems if they are characterized by the same pair of functions $(f,h)$.

\bdefin{onesideSCsystem}{(\bf The one-sided system} (cf. Def. 10 in \cite{yedla2012simple})
    The one-sided system is defined by the recursion system:
    \bear{onesideSC}
        &&p_\ell^{(i+1)}=\frac{1}{w}\sum_{\hat{\ell}=\max[1,\ell-w+1]}^\ell f\bigg(\nonumber\\
        &&\frac{1}{w}\sum_{\tilde \ell=\hat\ell}^{\min[\hat\ell+w-1,L]} h(p_{\tilde\ell}^{(i)});G\bigg),\nonumber\\
    \eear
    where $\mathbf{p}\in[0,1]^L$. Conditions (P1)-(P4) in \rdef{scalar_admissible_system} should be satisfied.

    The vector recursion form of \req{onesideSC} is given by:
    \beq{onesideSCvector}
        \mathbf{p}^{(i+1)}=\mathbf{A}^T\mathbf{f}(\mathbf{A}\mathbf{h}(\mathbf{p}^{(i)});G),
    \eeq
    where $\mathbf{A}$ is an $L\times L$ matrix defined as:
    \beq{matrixA}
        \mathbf{A}=\frac{1}{w}\begin{bmatrix}
            1      & 1      & \ldots  & 1      & 0      & \ldots  & 0 \\
            0      & 1      & 1      & \ldots  & 1      & \ddots & \vdots \\
            \vdots & \ddots & \ddots & \ddots & \ddots & \ddots & 0 \\
            0      & \ldots  & 0      & 1      & 1      & \ldots  & 1 \\
            0      & 0      & \ldots  & 0      & 1      & \ldots  & 1 \\
            0      & 0      & \ldots  & 0      & 0      & 1      & \vdots \\
            0      & 0      & \ldots  & 0      & 0      & 0      & 1
        \end{bmatrix},
    \eeq
    where the 1st, 2nd, \ldots, $(\tilde{L}-w+1)$-th rows of $\mathbf{A}$ contain $w$ ones.
\edefin

Next, we give two lemmas for the one-sided system. The first one illustrates that the one-sided system gives a component-wise upper bound of the basic spatially-coupled system with a boundary condition $s_\ell^{(i)}=s_{\tilde{L}}^{(i)}$ for $\ell=\tilde{L},\tilde{L}+1,\ldots, L$ after each iteration. The second one shows that $\mathbf{p}^{(i)}$ in \req{onesideSCvector} is a non-decreasing vector.

\blem{basic_sc_upperbounds}
    Consider a column vector $\mathbf{p}^{(i)}=(p^{(i)}_1,p^{(i)}_2, \ldots, p^{(i)}_{\tilde{L}})$ and a basic spatially-coupled system
    $$\mathbf{p}^{(i+1)}=\mathbf{A}_2\mathbf{f}(\mathbf{A}_2^T\mathbf{h}(\mathbf{p}^{(i)});G)$$
    with the initial condition $\mathbf{p}^{(0)}=(1, 1, \ldots, 1)$. Let $\mathbf{s}^{(i)}$ be a vector of length $L$ defined by the one-sided system
    $$\mathbf{s}^{(i+1)}=\mathbf{A}^T\mathbf{f}(\mathbf{A}\mathbf{h}(\mathbf{s}^{(i)});G)$$
    with the initial condition
    \beq{initssss}
        \mathbf{s}^{(0)}=(1,1,\ldots,1).
    \eeq
    If we enforce $s_\ell^{(i)}=s_{\tilde{L}}^{(i)}$ for $\ell=\tilde{L},\tilde{L}+1,\ldots,L$ after each iteration, we have
    \beq{upperbound}
        p^{(i)}_{\ell}\leq s^{(i)}_{\tilde{L}+w-\ell},\quad\forall \ell=1,2,\ldots,\tilde{L}.
    \eeq
    Thus, the percolation threshold of the basic spatially-coupled system is not smaller than that of the one-sided system.
\elem

\bproof
We prove this lemma by induction. The initial condition \req{initssss} shows that $p_\ell^{(0)}=s_{\tilde{L}+w-\ell}^{(0)}=1$ for $\ell=1,\ldots,\tilde{L}$. Thus, \req{upperbound} holds for $i = 0$. Next, by (P1), (P2), and the induction hypothesis, we have
\bear{upperbound1111}
    p_{\ell}^{(i+1)}&&=\frac{1}{w}\sum_{\hat{\ell}=\tilde{L}+1-\ell}^{\tilde{L}+w-\ell}f\bigg(\frac{1}{w}\sum_{\tilde \ell=\max[1,\tilde{L} - \hat{\ell} + 1]}^{\min[\tilde{L},\tilde{L} - \hat{\ell} + w]}h(p_{\tilde \ell}^{(i)});G\bigg)\nonumber\\
    &&\leq\frac{1}{w}\sum_{\hat{\ell}=\tilde{L}+1-\ell}^{\tilde{L}+w-\ell}f\bigg(\frac{1}{w}\sum_{\tilde \ell=\max[1,\tilde{L} - \hat{\ell} + 1]}^{\min[\tilde{L},\tilde{L} - \hat{\ell} + w]}h(s_{\tilde{L}+w-\tilde \ell}^{(i)});G\bigg).\nonumber\\
\eear

If we change the subscript by letting $\tilde \ell^\prime=\tilde{L}+w-\tilde \ell$, then the upper bound of \req{upperbound1111} could be further evaluated
\bear{upperbound2222}
    p_{\ell}^{(i+1)}&&\leq\frac{1}{w}\sum_{\hat{\ell}=\tilde{L}+1-\ell}^{\tilde{L}+w-\ell}f\bigg(\frac{1}{w} \sum_{\tilde \ell^\prime=\max[\hat\ell,w]}^{\min[\hat\ell+w-1,L]}h(s_{\tilde \ell^\prime}^{(i)});G\bigg)\nonumber\\
    &&=\frac{1}{w}\sum_{\hat{\ell}=\tilde{L}+1-\ell}^{\tilde{L}+w-\ell}f\bigg(\frac{1}{w} \sum_{\tilde \ell=\max[\hat\ell,w]}^{\min[\hat\ell+w-1,L]}h(s_{\tilde \ell}^{(i)});G\bigg)\nonumber\\
    &&\leq\frac{1}{w}\sum_{\hat{\ell}=\tilde{L}+1-\ell}^{\tilde{L}+w-\ell} f\bigg(\frac{1}{w}\sum_{\tilde \ell=\hat\ell}^{\min[\hat\ell+w-1,L]} h(s_{\tilde \ell}^{(i)});G\bigg).\nonumber\\
    &&=s_{\tilde{L}+w-\ell}^{(i+1)}.\nonumber\\
\eear

The first equality holds since we change the indices of the second summation. The second inequality follows from (P1) and the fact that more terms are included. The last equality follows from \req{onesideSC}.

Hence, for $\ell=1,2,\ldots,\tilde{L}$, \req{upperbound} also holds for the $(i+1)$-th iteration. This completes the proof.
\eproof

\rlem{basic_sc_upperbounds} implies that once the saturation theorem holds for the one-sided system, it automatically holds for the basic spatially-coupled system as well.

\blem{nondecreasing}
    Consider the one-sided system
    $$\mathbf{p}^{(i+1)}=\mathbf{A}^T\mathbf{f}(\mathbf{A}\mathbf{h}(\mathbf{p}^{(i)});G)$$
    with the initial condition
    \beq{initpppp}
        \mathbf{p}^{(0)}=(1,1,\ldots,1).
    \eeq
    If we enforce $p_\ell^{(i)}=p_{\tilde{L}}^{(i)}$ for $\ell=\tilde{L},\tilde{L}+1,\ldots,L$ after each iteration, then the vector $\mathbf{p}^{(i)}$ is non-decreasing, say
    \beq{orderpppp}
        p^{(i)}_0\leq p^{(i)}_1 \leq p^{(i)}_2 \leq\ldots\leq p^{(i)}_{\tilde{L}}=p^{(i)}_{\tilde{L}+1}=\ldots=p^{(i)}_{L}
    \eeq
    for all $i$.
\elem

\bproof
For $i = 0$, \req{orderpppp} holds by the initial condition \req{initpppp}. Suppose that \req{orderpppp} holds for the $i$-th iteration, we shall examine \req{orderpppp} for the $(i+1)$-th iteration. First, by \req{onesideSC}, we have
\bear{property1111}
    &&p_{\ell + 1}^{(i+1)} - p_{\ell}^{(i+1)} \nonumber\\
    &&= \frac{1}{w}\bigg( f\bigg(\frac{1}{w} \sum_{\tilde \ell = \ell + 1}^{\min{[L , \ell + w]}}h(p_{\tilde \ell}^{(i)}) ; G\bigg) \nonumber\\
    &&- f\bigg(\frac{1}{w} \sum_{\tilde \ell = \ell - w + 1}^{\min[\ell,L]}h(p_{\tilde \ell}^{(i)}) ; G\bigg)\bigg). \nonumber\\
\eear

\noindent Case I. $\ell < w$: By (P1) and (P2)
\bear{property2222}
    p_{\ell + 1}^{(i+1)} - p_{\ell}^{(i+1)}= \frac{1}{w}f\bigg(\frac{1}{w} \sum_{\tilde \ell = \ell + 1}^{\ell + w}h(p_{\tilde \ell}^{(i)}) ; G\bigg)>0. \nonumber\\
\eear

\noindent Case II. $w<\ell<\tilde{L}$:
\bear{property3333}
    &&p_{\ell + 1}^{(i+1)} - p_{\ell}^{(i+1)} \nonumber\\
    &&= \frac{1}{w}\bigg( f\bigg(\frac{1}{w} \sum_{\tilde \ell = \ell + 1}^{\ell + w}h(p_{\tilde \ell}^{(i)}) ; G\bigg) \nonumber\\
    &&- f\bigg(\frac{1}{w} \sum_{\tilde \ell = \ell - w + 1}^{\ell}h(p_{\tilde \ell}^{(i)}) ; G\bigg)\bigg) \nonumber\\
    &&>0. \nonumber\\
\eear
The inequality follows from (P1), (P2), and the induction hypothesis.

\noindent Case III. $\ell\geq \tilde{L}$: By the enforcement we make, $p_{\ell + 1}^{(i+1)} - p_{\ell}^{(i+1)}=0$

Thus, \req{orderpppp} holds for all $i$.
\eproof

Following Lemma 2 in \cite{yedla2012simple} and Lemma 14 in \cite{kudekar2011threshold}, we introduce a lemma that serves as the convolutional $(T, G, \Lambda(x), R, F,w)$-CPR version of Lemma 5 in \cite{chang2020Poisson}.

\blem{fixedpoint}
    Consider the recursion equation \req{onesideSC} with the condition that $p_\ell^{(i)}=p_{\tilde{L}}^{(i)}$ is enforced for $\ell=\tilde{L},\tilde{L}+1,\ldots,L$ after each iteration. If $f$ and $h$ is given by \req{fh}, then $p_\ell^{(i+1)}\leq p_\ell^{(i)}$ for all $\ell=1,\ldots{L}$ and all positive integer $i$. Also, \req{onesideSC} converges to a well-defined fixed point, say $$\lim_{i\to\infty}p^{(i)}:=p^{(\infty)}=(p^{(\infty)}_1,\ldots,p^{(\infty)}_{L}).$$
    Moreover, $p^{(\infty)}$ represents the (element-wise) largest solution among all solutions in $[0,1]^{L}$. In other words, if $\hat{p}=(\hat{p}_1,\ldots,\hat{p}_{L})$ is another solution, then $\hat{p}_\ell=p^{(\infty)}_\ell$, for $\ell=1,\ldots,{L}$. Furthermore, as a result of this and \rlem{basic_sc_upperbounds}, the density evolution equation \req{reversed2222} of CCPRs also converges to a well-defined fixed point.
\elem

\bproof
We begin by showing that $p_\ell^{(i)}$ is a decreasing sequence. Since $p_\ell^{(0)}=1$, then
\begin{equation*}
\begin{aligned}
    p_\ell^{(1)}&=\frac{1}{w}\sum_{\hat{\ell}=\max[1,\ell-w+1]}^\ell\bigg(1-\\
    &\quad P_\text{suc}\bigg(\frac{G\Lambda'(1)}{w}\sum_{\tilde \ell=\hat\ell}^{\min[\hat\ell+w-1,{L}]}\lambda(1)\bigg)\bigg)\\
    &\leq\frac{1}{w}\sum_{\hat{\ell}=\max[1,\ell-w+1]}^\ell(1-P_\text{suc}(G\Lambda'(1)))\\
    &=1-P_\text{suc}(G\Lambda'(1))\\
    &<p_\ell^{(0)}=1.
\end{aligned}
\end{equation*}

Suppose that $p_{\ell}^{(i+1)}<p_{\ell}^{(i)}$, by (P2) and that $P_\text{suc}$ is decreasing,
\begin{equation*}
\begin{aligned}
    &P_\text{suc}\bigg(\frac{G\Lambda^\prime(1)}{w}\sum_{\tilde \ell=\hat\ell}^{\min[\hat\ell+w-1,{L}]}\lambda(p_{\tilde\ell}^{(i+1)})\bigg)\\
    &>P_\text{suc}\bigg(\frac{G\Lambda^\prime(1)}{w}\sum_{\tilde\ell=\hat\ell}^{\min[\hat\ell+w-1,{L}]}\lambda(p_{\tilde\ell}^{(i)})\bigg),
\end{aligned}
\end{equation*}
for all $\hat\ell=1,\ldots,{L}$.

Multiplying the above inequality by $-1$ and adding 1 to each side, then taking the summation, we have
\begin{equation*}
\begin{aligned}
    &\sum_{\hat{\ell}=\max[1,\ell-w+1]}^\ell\bigg(1-P_\text{suc}\bigg(\frac{G\Lambda^\prime(1)}{w}\sum_{\tilde\ell=\hat\ell}^{\min[\hat\ell+w-1,{L}]}\lambda(p_{\tilde\ell}^{(i+1)})\bigg)\bigg)\\
    &<\sum_{\hat{\ell}=\max[1,\ell-w+1]}^\ell\bigg(1-P_\text{suc}\bigg(\frac{G\Lambda^\prime(1)}{w}\sum_{\tilde\ell=\hat\ell}^{\min[\hat\ell+w-1,{L}]}\lambda(p_{\tilde\ell}^{(i)})\bigg)\bigg).
\end{aligned}
\end{equation*}

Multiplying by $1/w$ gives that $p_\ell^{(i+2)}<p_\ell^{(i+1)}$. Hence, by induction, $p_\ell^{(i)}$ is a decreasing sequence. Moreover, since $p_\ell^{(i)}$ itself is a probability, it is bounded below by 0. Hence $p^{(\infty)}$ exists and is the fixed point of \req{onesideSC}.

Suppose $\hat{p}=(\hat{p}_1,\ldots,\hat{p}_{L})$ is another solution of \req{onesideSC}. It is obvious that $\hat{p}\leq 1=p^{(0)}_\ell$ for all $\ell$. Suppose $\hat{p}_\ell\leq p_\ell^{(i)}$ for some $i\geq1$, using the fact that $\lambda$ is increasing and the assumption that $P_\text{suc}$ is decreasing, from \req{onesideSC}, we have
\begin{equation*}
\begin{aligned}
    \hat{p}_\ell&=\frac{1}{w}\sum_{\hat{\ell}=\max[1,\ell-w+1]}^\ell\bigg(1\\
    &-P_\text{suc}\bigg(\frac{G\Lambda^\prime(1)}{w}\sum_{\tilde\ell=\hat\ell}^{\min[\hat\ell+w-1,{L}]}\lambda(\hat{p}_\ell)\bigg)\bigg)\\
    &\leq\frac{1}{w}\sum_{\hat{\ell}=\max[1,\ell-w+1]}^\ell\bigg(1\\
    &-P_\text{suc}\bigg(\frac{G\Lambda^\prime(1)}{w}\sum_{\tilde\ell=\hat\ell}^{\min[\hat\ell+w-1,{L}]}\lambda(p_{\tilde\ell}^{(i)})\bigg)\bigg)\\
    &=p^{(i+1)}_\ell
\end{aligned}
\end{equation*}

Hence, $\hat{p}\leq p_\ell^{(i)},\forall i$. Taking $i\to\infty$, we have $\hat{p}_\ell\leq p^{(\infty)}_\ell$ for all $\ell$.
\eproof

The next lemma further illustrates the proposition of the fixed point of the one-sided system.
\blem{no_fixed_point_uG}
    Let $\mathbf{p}=(p_1,p_2,\ldots,p_{L})$ be a fixed point of \req{onesideSC}. Suppose $\mathbf{p}\neq\mathbf{0}$ for $\ell=1,2,\ldots,L$, then the system has no fixed point with $p_L<u(G)$.
\elem

\bproof
By \rlem{nondecreasing} and \rlem{fixedpoint}, such fixed point is non-decreasing, say $p_1\leq p_2\leq\ldots\leq p_{L}$. Thus, \req{onesideSC} gives that,
\bear{est0000}
    p_L&&=\frac{1}{w}\sum_{\hat{\ell}=L-w+1}^L f\bigg(\frac{1}{w}\sum_{\tilde\ell=\hat\ell}^{\min[\hat\ell+w-1,L]} h(p_{\tilde\ell});G\bigg)\nonumber\\
    &&\leq \frac{1}{w} \sum_{\hat{\ell}=L-w+1}^L f\bigg(\frac{1}{w}\cdot w\cdot h(p_L);G\bigg)\nonumber\\
    &&=\frac{1}{w}\cdot w\cdot f(h(p_L);G)
\eear
Hence, $f(h(p_L);G)-p_L\geq0$. By \req{CCPR_potential_prime}, we have that
\bear{est1111}
    U^\prime(p_L;G)=\lambda^\prime (p_L)(p_L-1+P_\text{suc}(G\Lambda^\prime(p_L))\leq0.
\eear
Thus, by \rdef{minimum_unstable}, $p_L>u(G)$.
\eproof

Additionally, similar to the {potential function of the scalar admissible system $(f,h)$ in \req{potential_int_scalar}, we define the potential function of the one-sided system below.

\bdefin{potential}{(\bf The potential function of one-sided systems} (cf. Def. 11 in \cite{yedla2012simple})
    The potential function of the one-sided system describe in \rdef{onesideSCsystem} is the line integral along a curve $C$ in $\mathbb{R}^{\tilde{L}}$ joining $\mathbf{0}$ and $\mathbf{p}=(p_1,p_2,\ldots,p_{\tilde{L}})$,
    \beq{potential_int}
        U(\mathbf{p};G)=\int_C \mathbf{h}'(\mathbf{z})(\mathbf{z}-\mathbf{A}^T\mathbf{f}(\mathbf{A}\mathbf{h}(\mathbf{z});G))d\mathbf{z},\\
    \eeq
    where $\mathbf{h}'(\mathbf{p})=\text{diag}([h'(p_\ell)])$. Let
    $$H(\mathbf{p})=\int_C \mathbf{h}(\mathbf{z})d\mathbf{z}=\sum_{\ell=1}^LH(p_\ell)$$
    and
    $$F(\mathbf{p};G)=\int_C\mathbf{f}(\mathbf{p};G)d\mathbf{p}=\sum_{\ell=1}^LF(p_\ell;G).$$

    Then, \req{potential_int} could also be written in the form
    \beq{potential_int2}
        U(\mathbf{p};G)=\mathbf{h}(\mathbf{p})^T\mathbf{p}-H(\mathbf{p})-F(\mathbf{A}\mathbf{h}(\mathbf{p});G).
    \eeq
\edefin

The key insight is that the potential functions of the one-sided system and the scalar admissible system are connected through a shift operator. The advantage of this approach is that it eliminates the need for evaluating the line integral in \req{potential_int2} for exploring the properties of the stability region of the convolutional $(T, G, \Lambda(x), R, F,w)$-CPRs.

\blem{comb_lemma} (Lemma 3, in \cite{yedla2012simple})
    Let $\mathbf{p}=(p_1,\ldots,p_{L})\in[0,1]^{L}$ be a non-decreasing vector generated by averaging $\mathbf{q}\in[0,1]^{L}$ over a sliding window of size $w$. Let the shift operator $\mathbf{S}:\mathbb{R}^{L}\to\mathbb{R}^{L}$ be defined by $[\mathbf{S}\mathbf{p}]=(0,p_1,p_2,\ldots,p_{{L}-1})$. Then, $$||\mathbf{S}\mathbf{p}-\mathbf{p}||_\infty<\frac{1}{w}$$ and $$||\mathbf{S}\mathbf{p}-\mathbf{p}||_1=p_{L}=||\mathbf{p}||_\infty.$$
\elem

\bproof
Since
$$\mathbf{S}\mathbf{p}-\mathbf{p}=(-p_1,p_1-p_2,p_2-p_3,\ldots,p_{L-1}-p_{L}),$$
and $\mathbf{p}$ is non-decreasing,
$$||\mathbf{S}\mathbf{p}-\mathbf{p}||_1=|p_1|+\sum_{\ell=2}^{L} |p_{\ell-1}-p_{\ell}|=p_{L}=||\mathbf{p}||_\infty.$$
Furthermore,
\begin{equation*}
\begin{aligned}
    |p_\ell-p_{\ell-1}|&=\bigg|\frac{1}{w}\sum_{\hat\ell=0}^{w-1}(q_{\ell-\hat\ell}-q_{\ell-\hat\ell-1})\bigg|\\
    &=\frac{1}{w}|q_0-q_{L}|\\
    &\leq\frac{1}{w}
\end{aligned}
\end{equation*}
for any $\ell=2,\ldots,L$. So $||\mathbf{S}\mathbf{p}-\mathbf{p}||_\infty\leq\frac{1}{w}$.
\eproof

\blem{potentialchange} (cf. Lemma 4 in \cite{yedla2012simple})
    For the one-sided system defined in \req{onesideSC} and the potential defined in \rdef{potential}, a shift changes the potential by $U(\mathbf{S}\mathbf{p};G)-U(\mathbf{p};G)=-U(p_{L};G)$.
\elem

\bproof
First, we claim that
$$F([\mathbf{A}\mathbf{h}(\mathbf{S}\mathbf{p})]_\ell;G)=F([\mathbf{A}\mathbf{h}(\mathbf{p})]_{\ell-1};G).$$
This can be done by careful inspection: Since $\mathbf{h}(\mathbf{S}\mathbf{p})=(0,h(p_1),\ldots,h(p_{{L}-1}))$, by the definition of $\mathbf{A}$,
\begin{equation*}
\begin{aligned}
    \mathbf{A}\mathbf{h}(\mathbf{S}\mathbf{p})=\frac{1}{w}\bigg(
    &0,\\ 
    &h(p_1)\\ 
    &h(p_1)+h(p_2)\\ 
    &\ldots,\\
    &h(p_1)+h(p_2)+\ldots+h(p_{w-1}),\\ 
    &h(p_2)+h(p_3)\ldots+h(p_{w}),\\ 
    &\ldots,\\
    &h(p_{{L}+w-2})+\ldots+h(p_{{L}-1})\bigg). 
\end{aligned}
\end{equation*}

This proves the claim. Next, write the potential functions in the form of \req{potential_int2}, i.e.,
$$U(\mathbf{p};G)=\sum_{\ell=1}^{L}\big(h(p_\ell)p_\ell-H(p_\ell)-F([\mathbf{A}\mathbf{h}(\mathbf{p})]_\ell;G)\big),$$
and
\begin{equation*}
\begin{aligned}
    U(\mathbf{Sp};G)&=\sum_{\ell=1}^{L}\big(h(p_\ell)p_\ell-H(p_\ell)-F([\mathbf{A}\mathbf{h}(\mathbf{Sp})]_\ell;G)\big)\\
    &=\sum_{\ell=1}^{{L}-1}\big(h(p_\ell)p_\ell-H(p_\ell)\big)-\sum_{\ell=1}^{{L}}F([\mathbf{A}\mathbf{h}(\mathbf{Sp})]_\ell;G).
\end{aligned}
\end{equation*}

This gives that
\begin{equation*}
\begin{aligned}
    &U(\mathbf{Sp};G)-U(\mathbf{p};G)\\
    =&-h(p_{L})p_{L}+H(p_{L})\\
    &\quad+\sum_{\ell=1}^{{L}}\big(F([\mathbf{A}\mathbf{h}(\mathbf{Sp})]_\ell;G)-F([\mathbf{A}\mathbf{h}(\mathbf{p})]_\ell;G)\big)\\
    =&-h(p_{L})p_{L}+H(p_{L})+F(0;G)\\
    &\quad-F(h(p_{{L}+w-1})+\ldots+h(p_{{L}});G)\\
    =&-h(p_{L})p_{L}+H(p_{L})+F([\mathbf{A}\mathbf{h}(\mathbf{p})]_{L};G))\\
    =&-U(p_{L};G),
\end{aligned}
\end{equation*}
where that $F(0;G)$ comes from \req{potential3333}.
\eproof

\blem{Kfgupperbound} (cf. Lemma 5 in \cite{yedla2012simple})
    For the potential function of the one-sided system in \rdef{onesideSCsystem}, the norm of the Hessian $U''(\mathbf{p};G)$ is independent of $L$ and $w$. It satisfies
    $$||U''(\mathbf{p};G)||_\infty\leq K_{f,h}:=||h'||_\infty + ||h'||^2_\infty||f'||_\infty+||h''||_\infty,$$
    where
    $$||h||_\infty=\sup_{x\in[0,1]} |h(x)|$$ for functions $h:[0,1]\to\mathbb{R}$, and $$||\mathbf{A}||_\infty=\max_{1\leq i\leq {L}}\sum_{j=1}^{L}|a_{ij}|$$
    for the matrix $\mathbf{A}$.
\elem

\bproof
The Hessian is given by
\begin{equation*}
\begin{aligned}
    U''(\mathbf{p};G)&=\mathbf{h}'(\mathbf{p})-(\mathbf{A}\mathbf{h}'(\mathbf{p}))^T\mathbf{f}'(\mathbf{A}\mathbf{h}(\mathbf{p})G)\mathbf{A}\mathbf{h}'(\mathbf{p})\\
    &+\mathbf{h}''(\mathbf{p})\text{diag}(\mathbf{p}-\mathbf{A}^T\mathbf{f}(\mathbf{A}\mathbf{h}(\mathbf{p});G)).
\end{aligned}
\end{equation*}

Here, $\mathbf{h}''(x)=\text{diag}([h''(x_i)])$. Taking the norms,
\begin{equation*}
\begin{aligned}
    ||U''(\mathbf{p};G)||_\infty&=||\mathbf{h}'(\mathbf{p})||_\infty\\
    &+||\mathbf{A}\mathbf{h}'(\mathbf{p})||_\infty||\mathbf{f}'(\mathbf{A}\mathbf{h}(\mathbf{p})G)||_\infty||\mathbf{A}\mathbf{h}'(\mathbf{p})||_\infty\\
    &+||\mathbf{h}''(\mathbf{p})\text{diag}(\mathbf{p}-\mathbf{A}^T\mathbf{f}(\mathbf{A}\mathbf{h}(\mathbf{p});G))||_\infty.
\end{aligned}
\end{equation*}
Since $||\mathbf{A}||_\infty=1$, $||\mathbf{h}'(\mathbf{p})||_\infty\leq||h'||_\infty$, and all the elements inside $(\mathbf{p}-\mathbf{A}^T\mathbf{f}(\mathbf{A}\mathbf{h}(\mathbf{p});G))$ are in $[0,1]$. Hence,
$$||U''(\mathbf{p};G)||_\infty\leq K_{f,h}:=||h'||_\infty + ||h'||^2_\infty||f'||_\infty+||h''||_\infty.$$
\eproof

Now we prove \rthe{saturation_singleclass}.

\bproof
First, we claim that this theorem is true for the one-sided system \req{onesideSC} with the condition that $p_\ell^{(i)}=p_{\tilde{L}}^{(i)}$ is enforced for $\ell=\tilde{L},\tilde{L}+1,\ldots,L$ after each iteration. characterized by $(f,h)$ in \req{fh}.

Consider a load $G<G_{conv}^*$. This suffices to show that $\mathbf{0}$ is the only fixed point of \req{onesideSC}, given that the iteration times $i\to\infty$. Suppose that $\mathbf{p}\neq\mathbf{0}$ is a fixed point of \req{onesideSC}. By \rlem{no_fixed_point_uG}, $p_{L}>u(G)$. By \rlem{potentialchange}, expand $U(\mathbf{S}\mathbf{p};G)$ in a Taylor series around $U(\mathbf{p};G)$ with remainder.
\begin{equation*}
\begin{aligned}
    &U'(\mathbf{p};G)\cdot(\mathbf{S}\mathbf{p}-\mathbf{p})\\
    =&U(\mathbf{S}\mathbf{p};G)-U(\mathbf{p};G)\\
    &\quad -\int_0^1 (1-t)(\mathbf{S}\mathbf{p}-\mathbf{p})^TU^{\prime\prime}(\mathbf{p}(t);G)(\mathbf{S}\mathbf{p}-\mathbf{p})dt\\
    \leq&-U(p_{L};G)+\bigg|\int_0^1 (1-t)(\mathbf{S}\mathbf{p}-\mathbf{p})^TU^{\prime\prime}(\mathbf{p}(t);G)(\mathbf{S}\mathbf{p}-\mathbf{p})dt\bigg|\\
    \leq&-U(p_{L};G)+||\mathbf{S}\mathbf{p}-\mathbf{p}||_1||\mathbf{S}\mathbf{p}-\mathbf{p}||_\infty\max_{t\in[0,1]}||U''(\mathbf{p}(t);G)||_\infty\\
\end{aligned}
\end{equation*}

\rlem{nondecreasing} indicates that $\mathbf{p}$ is non-decreasing. Thus, we may put \rlem{comb_lemma} and \rlem{Kfgupperbound} into use. The last inequality could be further estimated:
$$U'(\mathbf{p};G)\cdot(\mathbf{S}\mathbf{p}-\mathbf{p})\leq-U(p_{L};G)+\frac{1}{w}\cdot p_{L}\cdot K_{f,h}.$$
Next, since $0\leq p_{L}\leq1$, using the condition $w>K_{f,h}/\Delta E(G)$ yields
\begin{equation*}
\begin{aligned}
    U'(\mathbf{p};G)\cdot(\mathbf{S}\mathbf{p}-\mathbf{p})&\leq-U(p_{L};G)+\frac{1}{w}K_{f,h}\\
    &\leq-U(p_{L};G)+\Delta E(G)\\
    &\leq 0.
\end{aligned}
\end{equation*}
The inequality follows from the definition of $\Delta E(G)$ and $p_L >u(G)$.

By \rlem{nondecreasing}, $\mathbf{p}$ is non-decreasing. Therefore, each component of $\mathbf{S}\mathbf{p}-\mathbf{p}$ is not greater than 0. Hence, there exists a component, say $\ell_0$-th of $U'(\mathbf{p};G)$, is greater than 0. Moreover, by (P2),
$$[U'(\mathbf{p};G)]_{\ell_0}=[h'(p_k)(\mathbf{p}-\mathbf{A}^T\mathbf{f}(\mathbf{A}\mathbf{h}(\mathbf{p});G))]_{\ell_0}$$
gives that
$$[\mathbf{A}^T\mathbf{f}(\mathbf{A}\mathbf{h}(\mathbf{p});G))]_{\ell_0}<p_{\ell_0}.$$

This shows that one more iteration reduces the value of $p_{\ell_0}$, for some $\ell_0$ in $1,\ldots,L$, and it contradicts with such $\mathbf{p}\neq\mathbf{0}$ is a fix point. Therefore, the only fixed point of \req{onesideSC} is $\mathbf{p}=\mathbf{0}$ under the load $G<G_{conv}^*$. This completes the proof of the claim.

Since the CCPR system with $K=J=1$ governed by \req{reversed2222} is the basic spatially-coupled system by \rlem{estimate}, \rlem{fixedpoint} demonstrates that \req{reversed2222} also converges to a well-defined fixed point. Moreover, since \rlem{basic_sc_upperbounds} shows that $\mathbf{p}$ element-wisely upper bounds the fixed point of \req{reversed2222}, \req{reversed2222} converges to $\mathbf{0}$ as $i\to\infty$. Hence, by \req{key2222}, \rthe{stable}, and \rthe{region}, all the loads $G<G_{conv}^*$ are stable for \req{reversed2222}. Finally, as \req{reversed2222} is a reversed-index version of \req{simplified}, all the loads $G<G_{conv}^*$ are stable for \req{simplified}.
\eproof

\section{List of notations}\label{notation_list}

We provide a list of notations used in this paper on the next page.

\clearpage

\topcaption{List of Notations}\label{notation}
\begin{supertabular}{|c|l|}
\hline
                $\mathbf{A}$                        & An $L\times L$ matrix defined in \req{matrixA}\\
                $\mathbf{A}_2$                      & An $L\times\tilde{L}$ matrix defined in \req{matrixA2}\\
                $a_{ij}$                            & The elements of the matrix $\mathbf{A}$\\
                $B$                                 & The bound of a capacity envelope\\
                $(b_1,\ldots,b_K)$                     & The capacity envelope\\
			$D$                                 & The maximum number of packets that can \\
                                                    & be successfully received in $D$-fold ALOHA\\
                $d$                                 & The degree of the regular degree distribution\\
                $\ex$                               & The expectation operator\\
                $e$                                 & Euler's number\\
                $F$                                 & The partition vector $F=(F_1,\ldots,F_J)$\\
                $F_j$                               & The fraction of Poisson receivers assigned to \\
                                                    & class $j$\\
                $F(p;G)$                            & The integral of $f(p;G)$ w.r.t. its first variable\\
                $F(\mathbf{p};G)$                   & The integral of $f(\mathbf{p};G)$ w.r.t. its first variable\\
                $f$                                 & One of the two functions that parameterize a \\
                                                    & one-sided system or a scalar admissible \\
                                                    & system. $f$ is a real-valued function with \\
                                                    & two variables\\
                $f^\prime$                          & The partial derivative of $f$ w.r.t. its first \\
                                                    & variable\\
                $\mathbf{f}(\mathbf{p};G)$          & A vector with $L$ components. \\
                                                    & $\mathbf{f}(\mathbf{p};G)=(f(p_1;G),\ldots,f(p_L;G))$\\
			$G$                                 & A vector of normalized offered load, \\
                                                    & $G=(G_1, G_2, \ldots, G_{K})$ \\
                $G_{conv}^*$                        & The potential threshold\\
                $G_{conv}^*(L,w)$                   & The percolation threshold of a CCPR with $L$ \\
                                                    & stages and window size $w$\\
                $G_{up}^*$                          & The solution to the equation $U(1;G)=0$\\
			$G_k$                               & The normalized offered load of class $k$\\
                $G^{(\ell)}$                        & The normalized offered load vector of $\ell^{th}$ \\
                                                    & stage, $G^{(\ell)}=(G^{(\ell)}_1,\ldots,G^{(\ell)}_K)$\\
			$G_k^{(\ell)}$                      & The normalized offered load of class $k$ on \\
                                                    & $\ell^{th}$ stage\\
                $G_s^*$                             & The single-system threshold\\
                $H(p)$                              & The integral of $h(p)$\\
                $H(\mathbf{p})$                     & The integral of $h(\mathbf{p})$\\
                $h$                                 & One of the two functions that parameterize a\\
                                                    & one-sided system or a scalar admissible \\
                                                    & system. $h$ is a real-valued function \\
                $\mathbf{h}(\mathbf{p})$            & A vector with $L$ components. \\
                                                    & $\mathbf{h}(\mathbf{p})=(h(p_1),\ldots,h(p_L))$ with one variable\\
			$i$                                 & The number of SIC iteration\\
                $J$                                 & The number of classes of receivers \\
                $j$                                 & The class of a receiver or the index of stages\\
                $K$                                 & The number of classes of users \\
                $k$                                 & The class of a packet or user\\
                $K_{f,h}$                           & A number defined in \rlem{Kfgupperbound} related to \\
                                                    & $f$ and $h$\\
                $L$                                 & The number of stages of CCPRs\\
                $\tilde{L}$                         & $L-w+1$\\
			$L_k$                               & The number of copies of a class $k$ packet \\
                $\ell$                              & The index of stages of a CCPR\\
                $m(j)$                              & $m(j)=\min\{1,\hat{\ell}\ominus(w-1)\}$\\
                $n$                                 & The deterministic load $n=(n_1,n_2,\ldots,n_K)$\\
                $P_{\text{suc}}(\rho)$              & The probability that a packet is \\
                                                    & successfully received if the receiver is\\
                                                    & subject to a Poisson offered load $\rho$\\
                $P_{\text{suc}}^{-1}$               & The inverse of $P_{\text{suc}}$\\
                $P_{\text{suc},D}(\rho)$            & The success function of the $D$-fold \\
                                                    & ALOHA system\\

                $P_{\text{suc},k}(\rho)$            & The probability that a class $k$ packet is \\
                                                    & successfully received if the receiver is \\
                                                    & subject to a Poisson offered load $\rho$\\
			$P_{\text{suc},k,j}(\rho)$          & The probability that a class $k$ packet is \\
                                                    & successfully received by a class $j$ \\
                                                    & receiver if the receiver is subject to a \\
                                                    & Poisson offered load $\rho$\\
                $\tilde P_{{\rm suc},k}^{(i)}$      & The probability that a packet sent from \\
                                                    & a randomly selected class $k$ user can \\
                                                    & be successfully received after the \\
                                                    & $i^{th}$ iteration\\
                $\tilde P_{{\rm suc},k,\ell}^{(i)}$ & The probability that a packet sent from \\
                                                    & a randomly selected class $k$ user in the \\
                                                    & $\ell^{th}$ stage can be successfully \\
                                                    & received after the $i^{th}$ iteration \\
                $\mathbf{p}$                        & A vector. $\mathbf{p}=(p_1,\ldots,p_L)$\\
                $\tilde{p}_\ell^{(i)}$              & The success probability after reversing the \\
                                                    & indices of \req{simplified}. $\tilde{p}_\ell^{(i)}=p_{{\tilde{L}}-\ell+1}^{(i)}$\\
                $p^{(0)}$                           & The initial vector of $(p_{1}^{(i)},\ldots,p_{L}^{(i)})$ \\
                $p^{(i)}$                           & The probability that the receiver end of \\
                                                    & a randomly selected edge has not been \\
                                                    & successfully received after the $i^{th}$ \\
                                                    & SIC iteration \\
                $p^{(\infty)}$                      & The limiting vector of $(p_{1}^{(i)},\ldots,p_{L}^{(i)})$ \\
                                                    & with the initial vector $p^{(0)}={\bf 1}$ \\
			$p_k^{(i)}$                         & The probability that the receiver end of \\
                                                    & a randomly selected class $k$ edge has not \\
                                                    & been successfully received after the \\
                                                    & $i^{th}$ SIC iteration \\
                $p_{\ell}^{(i)}$                    & The probability that the receiver end of a \\
                                                    & randomly selected edge in the $\ell^{th}$ \\
                                                    & stage has not been successfully received \\
                                                    & after the $i^{th}$ SIC iteration\\
			$p_{k,j}^{(i)}$                     & The probability that the receiver end of \\
                                                    & a randomly selected class $(k,j)$-edge has \\
                                                    & not been successfully received after the \\
                                                    & $i^{th}$ SIC iteration\\
                $p_{k,\ell}^{(i)}$                  & The probability that the receiver end of \\
                                                    & a randomly selected class $k$ edge in $\ell^{th}$ \\
                                                    & stage has not been successfully received \\
                                                    & after the $i^{th}$ SIC iteration\\

                $p_{k,j,\ell}^{(i)}$                & The probability that the receiver end of a randomly selected class $(k,j)$-edge in the $\ell^{th}$ stage has not\\
                                                    & been successfully received after the $i^{th}$ SIC iteration\\
                $q^{(i)}$                           & A vector of $q_k^{(i)}$, $q^{(i)}=(q_{1}^{(i)}, q_{2}^{(i)}, \ldots, q_{K}^{(i)})$\\
                $q^{(0)}$                           & The initial vector of $q^{(i)}$ \\
                $q^{(\infty)}$                      & The limiting vector of $q^{(i)}$ with the initial vector $q^{(0)}={\bf 1}$ \\
			$q_k^{(i)}$                         & The probability that the user end of a randomly selected class $k$ edge has not been successfully received \\
                                                    & after the $i^{th}$ SIC iteration\\
                $q_\ell^{(i)}$                      & A vector of $q_{k,\ell}^{(i)}$, $q_\ell^{(i)}=(q_{1,\ell}^{(i)}, q_{2,\ell}^{(i)}, \ldots, q_{K,\ell}^{(i)})$\\
                $q_{k,\ell}^{(i)}$                  & The probability that the user end of a randomly selected class $k$ edge in the $\ell^{th}$ stage has not been \\
                                                    & successfully received after the $i^{th}$ SIC iteration\\
                $R$                                 & The $K\times J$ routing matrix $R=(r_{k,j})$\\
                $\mathbb{R}$                        & The set of real numbers\\
			$\Rj$                               & A vector of parameters for class $j$ receiver, $\Rj=(\frac{r_{1,j}}{F_j}, \frac{r_{2,j}}{F_j}, \ldots, \frac{r_{K,j}}{F_j})$\\
			$r_{k,j}$                           & The routing probability that a class $k$ packet transmitted to a class $j$ receiver \\
                $r_j$                               & $r_j=r_{1,j}+\ldots+r_{K,j}$\\
                $S$                                 & The stability region or the capacity region\\
                $\mathbf{S}$                        & The shift operator\\
			$S_L$                               & The stability region of a CCPR with $L$ stages\\
			$T$                                 & The number of Poisson receivers\\
                $U(p;G)$                            & The potential function of a scalar admissible system\\
                $U(\mathbf{p};G)$                   & The potential function of a one-sided system\\
                $u(G)$                              & The minimum unstable fixed point\\
                $w$                                 & The smooth window size of CCPRs\\
                $X_k(t)$                            & The number of class $k$ packets sent to the $t^{th}$ receiver \\
                $Y_k(t)$                            & The number of class $k$ packets that are  actually decoded by the $t^{th}$ receiver \\
                ${\cal Z}^+$                        & The set of nonnegative integers\\
                $\Delta E(G)$                       & The energy gap\\
                $\delta$                            & The step size\\
                $\Theta_k$                          & The throughput for a $(P_{\text{suc},1}(\rho),\ldots,P_{\text{suc},K}(\rho))$-Poisson receiver subject to a Poisson offered load $\rho$\\
			$\Lambda_{k,d}$                     & The probability that a class $k$ packet is transmitted $d$ times \\
                $\Lambda^\prime(x)$                 & A vector of $\Lambda_{k}^\prime(x)$, $\Lambda^\prime (x)=(\Lambda^\prime_1 (x), \Lambda^\prime_2 (x), \ldots, \Lambda^\prime_K (x))$\\
                $\Lambda_k(x)$                      & The generating function of the degree distribution of a class $k$ user \\
			$\Lambda_{k}^\prime(x)$             & The derivative of $\Lambda_k(x)$\\
			$\Lambda_{k}^\prime(1)$             & The mean degree of a class $k$ user node\\
			$\lambda_{k,d}$                     & The probability that the user end of a randomly selected class $k$ edge has additional $d$ edges excluding the \\
                                                    & randomly selected edge \\
        	$\lambda_k(x)$                      & The generating function of the excess degree distribution of a class $k$ user \\
                $\mu_j$                             & The mean of the Poisson random $\sum_{k=1}^K b_kX_k(t)$\\
			$\rho$                              & The Poisson offered load $\rho=(\rho_1, \ldots, \rho_K)$\\
			$\rho_k$                            & The Poisson offered load of class $k$\\
                $\trhoj$                            & The Poisson offered load at a class $j$ Poisson receiver $\trhoj=(\rho_{1,j}, \rho_{2,j}, \ldots, \rho_{K,j})$\\
			$\rho_{k,j}$                        & The Poisson offered load of class $k$ packets to a class $j$ Poisson receiver\\
                $\tilde{\rho}_{j,\ell}$             & The vector of $\rho_{k,j,\ell}$. $\tilde{\rho}_{j,\ell}=(\rho_{1,j,\ell},\rho_{2,j,\ell},\ldots,\rho_{K,j,\ell})$ \\
                $\rho_{k,j,\ell}$                   & The Poisson offered load of class $k$ packets to a class $j$ Poisson receiver in the $\ell^{th}$ stage\\
                $\phi$                              & The $\phi$-ALOHA receiver. $\phi(n)=(\phi_1(n),\ldots,\phi_K(n))$\\
\hline
\end{supertabular}

\end{document}